\documentclass[useAMS,usenatbib,usegraphicx,11pt]{article} 

\pdfoutput=1
\usepackage{jcappub}
\usepackage[english]{babel}
\usepackage{xcolor}
\usepackage{soul}
\usepackage{lineno} 
\usepackage{multirow}
\usepackage{colortbl}
\usepackage{units}
\usepackage{siunitx}
\usepackage{mathrsfs} 
\usepackage[super]{nth}
\usepackage{bm}
\usepackage[nameinlink,noabbrev]{cleveref}
\usepackage{subfigure}
\usepackage[hang, small, bf]{caption}
\usepackage{sidecap}
\usepackage{afterpage}

\newcommand{\gr}{$\gamma$-ray}
\newcommand{\grs}{$\gamma$-rays}

\newcommand{\eq}[1]{Eq.~(\ref{#1})}
\newcommand{\beq}{\begin{equation}}
\newcommand{\eeq}{\end{equation}}
\newcommand{\balign}{\begin{align}}
\newcommand{\ealign}{\end{align}}

\newcommand{\lcdm}{{\ifmmode \Lambda{\rm CDM} \else $\Lambda{\rm CDM}$\fi}}
\newcommand{\mchi}{\ensuremath{m_{\chi}}}
\newcommand{\Msol}{\ensuremath{\rm M_\odot}}

\newcommand{\fsub}{\ensuremath{f_{\rm subs}}}

\newcommand{\alpham}{\ensuremath{\alpha_m}}
\newcommand{\alphaM}{\ensuremath{\alpha_M}}

\newcommand{\clumpy}{{\tt CLUMPY}}

\newcommand{\sigmav}{\ensuremath{\langle \sigma v\rangle}}

\newcommand{\dd}{\ensuremath{\mathrm{d}}}
\newcommand{\fermi}{{\em Fermi}-LAT}

\newcommand{\planck}{{\it Planck}}

\crefname{subsection}{\S}{\S}
\crefname{equation}{Eq.}{Eq.}

\title{\boldmath Extragalactic diffuse $\gamma$-rays  from dark
matter annihilation: revised prediction and full modelling uncertainties}

\author[a,b,c]{M. H\"utten,}
\author[c]{C. Combet,}
\author[c]{D. Maurin.}
\affiliation[a]{Humboldt-Universit\"{a}t zu Berlin, Newtonstra{\ss}e 15, D-12489 Berlin, Germany}
\affiliation[b]{DESY, Platanenallee 6, D-15738 Zeuthen,  Germany}
\affiliation[c]{LPSC, Universit\'e Grenoble-Alpes, CNRS/IN2P3, 53 avenue des Martyrs, 38026 Grenoble, France}
\emailAdd{moritz.huetten@hu-berlin.de}
\emailAdd{celine.combet@lpsc.in2p3.fr}
\emailAdd{dmaurin@lpsc.in2p3.fr}

\abstract{
Recent high-energy data from \fermi{} on the diffuse {\gr} background have been used to set among the best constraints on annihilating TeV  cold dark matter candidates. In order to assess the robustness of these limits, we revisit and update the calculation of the isotropic extragalactic \gr{} intensity from dark matter annihilation. The emission from halos with masses $\geq10^{10}$~\Msol{} provides a robust lower bound on the predicted intensity. The intensity including smaller halos whose properties are extrapolated from their higher mass counterparts is typically 5 times higher, and  boost from subhalos yields an additional factor $\sim 1.5$. We also rank the uncertainties from all ingredients and provide a detailed error budget for them. Overall, our fiducial intensity is a factor 5 lower than the one  derived by the \fermi{} collaboration in their latest analysis. This indicates that the limits set on extragalactic dark matter annihilations could be relaxed by the same factor. We also calculate the expected intensity for self-interacting dark matter in massive halos and find  the emission reduced by a  factor 3 compared to the collisionless counterpart. The next release of the \clumpy{} code will provide all the tools necessary to reproduce and ease future improvements of this prediction.}


\begin{document}
\maketitle
\flushbottom


\section{Introduction}
\label{sec:intro}

The diffuse \gr{} background (DGRB) is, on angular scales larger than one degree, an isotropic radiation believed to be mostly of extragalactic origin. While it is established that several classes of astrophysical sources contribute to the DGRB (active galactic nuclei, star-forming galaxies, millisecond pulsars), its exact composition remains uncertain~\cite{2015PhR...598....1F}. The DGRB is also one of the many targets for indirect dark matter (DM) searches \cite{2012PDU.....1..194B,2015PhR...598....1F}, via annihilation or decay of DM at Galactic and cosmological scales.

Indirect signs of annihilating DM due to secondary $\gamma$-radiation were first considered at the end of the 1970s in the context of diffuse astrophysical \gr{} emissions~\cite{1978ApJ...223.1015G,1978ApJ...223.1032S}. It was concluded that the  \gr{} signal from extragalactic DM was negligible compared to the one from Galactic DM, as has been confirmed by several subsequent studies (e.g., \cite{1984PhRvL..53..624S,1991A&A...249....1G}). Twenty years later, the calculation was revisited \cite{2001PhRvL..87y1301B,2002PhRvD..66l3502U} based on an improved understanding of structure formation in the $\Lambda$CDM cosmological paradigm: contrarily to previous estimates, the DGRB was found to be a promising entity to probe for signatures of DM annihilation. Many efforts followed to refine this calculation (see \cite{2015PhR...598....1F} for a comprehensive list of references), or to go beyond the simple average calculation to increase the sensitivity to the DM signal against astrophysical backgrounds (via the photon distribution function \cite{2015JCAP...09..027F}, searching for a small-scale anisotropy in the DGRB via auto-correlations \cite{2006PhRvD..73b3521A} or cross-correlations with galaxy catalogues \cite{2008PhRvD..77l3518C}). Based on these calculations and the analysis of four years of \fermi{} data \cite{2010JCAP...04..014A,2015JCAP...09..008T}, several authors \cite{2015JCAP...09..008T,2016ApJ...819...44A,2016PhRvD..94l3005F,2017ChPhC..41d5104L} recently concluded that the limits on dark matter candidates derived from the DGRB are competitive with the best constraints on DM set by dwarf spheroidal galaxies \cite{2015PhRvL.115w1301A,2015ApJ...809L...4D}, and that they are currently the best limits set by \fermi{} data at TeV masses.

In consequence, it is crucial to understand and reduce as much as possible the modelling uncertainties when estimating this exotic extragalactic contribution to the DGRB. These uncertainties come from our limited knowledge of several input ingredients (mass function of the DM halos, their mass concentration and density profile, etc.), as already discussed in the literature. Most of these studies were performed in the framework of the halo model description in real space
\cite{2002PhRvD..66l3502U,2003MNRAS.339..505T,2005PhRvD..71b1303A,2010JCAP...04..014A,2010MNRAS.405..593Z,2014MNRAS.439.2728M,2014PhRvD..89h3001N,2015JCAP...09..008T,2016JCAP...08..069M},
with the exception of \cite{2012MNRAS.421L..87S,2014MNRAS.441.1861S} who proposed an approach based on the non-linear matter power spectrum in Fourier space. The two approaches are complementary, with slightly different uncertainties, and they were compared and found in reasonable agreement in \cite{2015JCAP...09..008T}. Overall, these authors argue for a factor 20 uncertainty on the cosmological-induced DM signal.

This work is based on halo model descriptions relying on recent results for the halo mass function \cite{2016MNRAS.457.4340K,2016MNRAS.462..893R} and the mass-concentration relation \cite{2015MNRAS.452.1217C,2016MNRAS.460.1214L} to provide an updated value for the cosmological signal from DM annihilations. Indeed, recent halo mass functions are based on updated cosmological parameters, more in line with the \planck{} cosmology \cite{2016A&A...594A..13P}; recent mass-concentration relations also better account for the connection between the halo accretion history and formation time, allowing for an improved description of the concentration, valid in principle up to any redshift. For comparison purposes, we also consider other descriptions \cite{2008ApJ...688..709T,2014MNRAS.442.2271S} that have been used in previous estimates of the extragalactic DM contribution to the DGRB.

Our main goal is to bracket more closely the \gr{} signal uncertainties and to rank the various sources of uncertainties. It is useful to split the \gr{} signal into (i) a robustly determined lower-limit contribution from high-mass halos only ($M\gtrsim \unit[10^{10}]{\Msol}$), and (ii) a larger but more uncertain contribution from less massive halos; the latter could actually be dominant in CDM scenarios in which very low-mass subhalos survive, but largely {bf negligible} for some classes of interacting DM (IDM). The many ingredients and calculations presented below will be available in a forthcoming release of the \clumpy{} code\footnote{\clumpy{} \cite{2012CoPhC.183..656C,2016CoPhC.200..336B} is a public code for the calculation of annihilating or decaying DM in various DM targets. It was previously used for dwarf spheroidal galaxies \cite{2011MNRAS.418.1526C,2015MNRAS.446.3002B,2015MNRAS.453..849B}, galaxy clusters \cite{2012PhRvD..85f3517C,2012MNRAS.425..477N}, and Galactic dark clumps \cite{2016JCAP...09..047H}. It is being developed for the extragalactic emissions (average, distribution function, and angular power spectrum).}. In the context of the continuous increase of \fermi{} data (see the sensitivity projections for DM targets \cite{2016PhR...636....1C}), it is important to be able to easily repeat and improve on the calculation, as soon as more robust ingredients become available. In particular, the absolute DM intensity level also determines the feasibility of photon statistic or spatial cross-correlation studies in search for DM.

The paper is organised as follows: in \cref{sec:basics}, we recall the formalism to calculate the contribution of extragalactic annihilating DM to the DGRB. The \gr{} signal in the context of collisionless CDM scenarios is presented in \cref{sec:flux_CDM}, where we discuss the lower limit of this contribution, its uncertainties, and the signal dependence on low-mass extrapolations of the mass function. We discuss the CDM results in \cref{sec:discussion}, also commenting on self-interacting DM (SIDM), before concluding in \cref{sec:conclusion}.


\section{Modelling the \gr{} emission from extragalactic DM halos}
\label{sec:basics}

The extragalactic differential \gr{} intensity of annihilating DM, averaged over the whole sky, is given by
\beq
\!\!\!\!I(E_\gamma)=\left\langle\frac{\dd  \Phi}{\dd
E_\gamma\,\dd\Omega}\right\rangle_{\rm \!\!sky} \!\!\!=
\frac{\overline{\varrho}^2_{\mathrm{DM},\,0}
\,\sigmav}{8\pi\,m_{\chi}^2} \int\limits_0^{z_\mathrm{max}} \!\!c\,\dd z\,
\frac{(1+z)^3}{H(z)}\left\langle\delta^2(z)\right\rangle\left.\frac{\dd N^{\gamma}_{\mathrm{source}}}{\dd E_{\mathrm e}}\right|_{E_{\mathrm e}=(1+z)E_\gamma} \!\!\!\!\!\times e^{-\tau(z,\, E_\gamma)}\,,
\label{eq:mean_intensity}
\eeq
with $E_\gamma$ the observed energy, $\Phi$ the flux, $\dd\Omega$ the elementary solid angle, $c$ the speed of light, $\overline{\varrho}_{\mathrm{DM},\,0}$ the DM density of the Universe today, and $H(z)$ the Hubble constant at redshift $z$. The remaining quantities in the equation are related to four important ingredients of the calculation, as described below.

\paragraph{Properties of the DM candidate.} A DM candidate is described by its mass $m_\chi$, the velocity averaged annihilation cross section $\sigmav$, and $\dd N^{\gamma}_{\mathrm{source}}/\dd E_{\mathrm e}$ the differential \gr{} yield per annihilation|the yield must be evaluated at $E_e=(1+z)E_\gamma$ to get a photon at $E_\gamma$ today. Throughout this paper, we assume a neutralino-like Majorana DM candidate.\footnote{For a Dirac particle, the factor $8\pi$ in the denominator of \cref{eq:mean_intensity} would be $16\pi$.} We rely on the \clumpy{} implementation \cite{2016CoPhC.200..336B} of PPPC4DMID \cite{2011JCAP...03..051C} to calculate the yield for various final states, e.g., $b\bar{b}$ quarks or $\tau^+\tau^-$ leptons, and only consider prompt \grs{} produced in the  hadronization and decay cascades after the DM particles' self-annihilation.  In principle, the source term should also account for inverse Compton (IC) upscattered photons of the Cosmic Microwave Background (CMB) off DM-induced electrons at relativistic energies $E_e$. These upscattered CMB photons have characteristic energies of $E_{\rm IC}\approx3.4\times10^{-2}\,(1+z)(E_e/(\rm 100~GeV))^2$~GeV \cite{2010PhRvD..81d3505B} and  their spectrum peaks at lower energies compared to the prompt \gr{} emission. The relative importance of IC emission w.r.t. the prompt \grs{}  and its energy dependence as a function of the final state and DM mass are illustrated in figure 1 of \cite{2010PhRvD..81d3505B}. In this study, we are mostly interested in the high-energy end of the \gr{} spectrum and the relative uncertainty on the signal, which is independent of the low-energy spectral shape. For this reason, we do not include the contribution from IC scattering here, but remind that it could change the DM \gr{} spectra shown in later plots at  energies  $E_\gamma\lesssim 10^{-2}\,\mchi$.

\paragraph{Transparency of the Universe to \grs{}.} The term $\tau(z,\,E_\gamma)$ in \cref{eq:mean_intensity} is the optical depth due to $e^+e^-$ pair production of \grs{} on the infrared and optical extragalactic background light (EBL) and on the cosmic microwave background (CMB). This encodes the increasing attenuation of the extragalactic \grs{} from GeV to TeV energies and beyond. The values of the optical depth at various $z$ and $E_\gamma$ and its overall impact on the \gr{} spectrum are further discussed in \cref{subsec:benchmark_model}.

\paragraph{Intensity multiplier from DM distribution.} Density fluctuations in the early Universe give rise to an inhomogeneous distribution of DM, $\varrho_{\mathrm{DM}}(\Omega,z) = \delta(\Omega,z)\times\overline{\varrho}_{\mathrm{DM}}(z)$. The inhomogeneous mass distribution boosts the rate of DM annihilations, expressed by the intensity multiplier $\langle\delta^2 \rangle=1 + \mathrm{Var}(\delta)$  in \cref{eq:mean_intensity}. For smoothly distributed DM, $\delta\equiv 1$, $\mathrm{Var}(\delta)=0$  and $\langle \delta^2 \rangle = 1$, whereas for a high density contrast, $\langle\delta^2 \rangle\approx \mathrm{Var}(\delta) \gg 1$. The contribution to the \gr{} intensity from the inhomogeneous Universe dominates at $z\lesssim 50$ \cite{2014MNRAS.439.2728M,2015MNRAS.452.1217C} and we only consider this contribution in this work (see also \cref{fig:Fluxmultiplier_integrated} in \cref{sec:intensity_multiplier_bricks}).

The quantity $\langle \delta^2(z) \rangle$ can either be directly computed from the nonlinear matter power spectrum $P_{\rm nl}(k,z)$ \cite{2012MNRAS.421L..87S,2014MNRAS.441.1861S}, or from using the halo model approach that we adopt here (as, e.g., followed by \cite{2002PhRvD..66l3502U,2013PhRvD..87l3539A,2016JCAP...08..069M}). In the halo model setup, the intensity multiplier from the inhomogeneous Universe is written as
\beq
\left\langle\delta^2(z)\right\rangle = \frac{1}{\overline{\varrho}^2_{\mathrm{m,0}}}\; \int \dd M \frac{\dd n}{\dd M}(M,z)\times \mathcal{L}(M,z) \,,
\label{eq:delta_dm_annihil}
\eeq
with today's mean total matter density $\overline{\varrho}_{\mathrm{m,\,0}}$, the halo mass function ${\dd n}/\dd M$ (i.e. the comoving number density of halos of mass $M$ at redshift $z$), and  the comoving one-halo luminosity, $\mathcal{L}(M,z)$; these terms are described in the following way:

\begin{itemize}
\item The halo mass function $ {\dd n}/{\dd M}$ is usually expressed in terms of the variance $\sigma$ of the density fluctuations in the linear regime and of the multiplicity function $f(\sigma,\,z)$ that encodes nonlinear structure formation,
\beq
\frac{\mathrm{d}n}{\mathrm{d}M}(M,\,z)  = f(\sigma,z)\; \frac{\overline{\varrho}_{\rm m,0}}{M}\, \frac{\mathrm{d}\ln \sigma^{-1}}{\mathrm{d}M}\,.
\label{eq:halomassfunction}
\eeq
The density variance $\sigma$ is a function of the linear matter power spectrum, $P_{\rm lin}(k,z=0)$, according to
\beq
\sigma^2(M,\,z) = \frac{D(z)^2}{2\pi^2} \int P_{\rm lin}(k, z=0)\, \widehat{W}^2(kR)\,k^2\,\mathrm{d}k\,,
\label{eq:sigma2}
\eeq
where a collapsing region of comoving radius $R=[3M/(4\pi\overline{\varrho}_{\rm m,0})]^{1/3}$ containing the mass $M$ is defined by a spherical top-hat window $W$, with $\widehat{W}(kR) = 3\,(kR)^{-3} \,[\sin(kR)  - kR \,\cos(kR)]$ in Fourier space. We use the  \texttt{CLASS} code \cite{2011arXiv1104.2932L} to compute $P_{\rm lin}(k,z=0)$  and we evolve $\sigma(M,\,z)^2$ to higher redshifts via the growth factor $D(z) = g(z)/g(z=0)$ with \cite{1977MNRAS.179..351H}
\beq
g(z) = \frac{5}{2}\, \Omega_{\rm m,0}\,H_0^{\;2}\times H(z)\,\int_z^\infty \frac{1+z'}{H^3(z')}\,\dd z'\,,
\eeq
where $\Omega_{\rm m,0} = \frac{8\pi G}{3}\,H_0^{\;-2}\times \overline{\varrho}_{\rm m,0}$ and $G$ is the gravitational constant.
\item The comoving one-halo luminosity,
\beq
\mathcal{L}(M_{\Delta},z) = \int\dd V\, \rho^2_{\mathrm{halo}} = 4\pi \int_0^{R_{\Delta}}\dd r \,r^2\, \rho^2_{\mathrm{halo}}\,,
\label{eq:luminosity}
\eeq
is a function of the halo mass $M_{\Delta}$, a given halo density profile\footnote{The
radius $r_{-2}$ is defined as $\dd \log \rho_{\mathrm{halo}}/\dd \log r|_{r=r_{-2}}=-2$, and $\rho_{-2} := \rho_{\mathrm{halo}}(r=r_{-2})$.} $\rho_{\mathrm{halo}}(r;\,\rho_{-2},\,r_{-2})$, and a mass-concentration relation
\begin{equation}
c_{\Delta}(M_{\Delta},z) := \frac{R_{\Delta}}{r_{-2}}
\label{eq:C-M}
\end{equation}
that is required to determine the normalisation of the profile given the halo mass. The subscript $\Delta$ denotes that a halo mass $M_{\Delta}$ is connected to its size, $R_{\Delta}$, via the relation
\beq
R_{\Delta}( M_{\Delta}, z) =\left( \frac{3\, M_{\Delta}}{ 4\pi \times \Delta(z) \times \varrho_{\rm c}(z)}\right)^{1/ 3} \times (1 + z)\,,
\label{eq:rdeltamdelta}
\eeq
where $\varrho_{\rm c}= \frac{3}{8\pi G} \,H^2(z)$ is the critical density of the Universe. While numerical simulations provide reasonably precise and scale-invariant density profiles $\rho_{\mathrm{halo}}$, the halo mass depends on the definition used for the halo outer bound \cite[e.g.,][]{2016MNRAS.456.2486D}.
 \end{itemize}

\paragraph{Redshift range.} In principle, all radiation from DM relic annihilation after the recombination era would contribute to today's observed intensity, \cref{eq:mean_intensity}. However, the cosmic \gr{} horizon for $E_{\gamma}\sim\unit[100]{GeV}$ photons due to pair-production on the EBL lies at $z\sim 1$ \cite{2013ApJ...770...77D} and at lower redshifts for higher energies. Conversely, at the lowest energies, $E_{\rm \gamma}\lesssim \unit[20]{GeV}$, the Universe is mostly transparent to \grs{} and radiation from high redshifts could significantly contribute  to the intensity. In the remainder of the paper,  we perform the numerical integration of the \gr{} intensity up to $z_{\rm max}=10$, corresponding to the highest redshift regime for which calculations of the DM distribution and models of the EBL are available.\footnote{Before the formation of the first stars, the only low-energy radiation background in the Universe was the CMB.}

\section{Exotic \gr{} intensity in the collisionless CDM paradigm}
\label{sec:flux_CDM}
Almost all ingredients above are related to DM halo properties, which are mostly studied by means of numerical simulations at cosmological scales \cite{2012PDU.....1...50K}. Their finite mass resolution does not generally allow the characterisation of the halo population below $M \sim \unit[10^{10}]{\Msol}$, while the smallest DM protohalos can possibly form down to $\unit[10^{-12}]{\Msol}$ \cite{2006PhRvL..97c1301P,2009NJPh...11j5027B,2014PhyU...57....1B}|the exact mass cut-off is related to the properties and kinetic decoupling of the DM candidate. This is an issue as this low-mass population may be responsible for a large part of the signal, and accounting for it is the largest source of uncertainty on the \gr{} signal (see \cref{subsec:lowmass_contrib}). Providing a universal description of the halo properties at all scales and all redshifts is further complicated by the various origins of the physics processes and environments in which these halos form and evolve \cite{2017ARA&A..55..343B}.

The halo properties used in our calculation are taken from the most advanced results from the literature, favouring those attempting to provide a universal description over all scales and epochs. The various results obtained by different groups allow us to define a range of values for these ingredients. Propagated to the intensity calculation, they provide reasonable and hopefully realistic uncertainties for the \gr{} emission estimate. The reference parameters and the range or configurations used are gathered in \cref{tab:benchmarkmodel}; the corresponding results and discussions are detailed in \cref{subsec:benchmark_model} and \cref{subsec:lowmass_contrib}.

\begin{table}[t]
\centering
\begin{tabular}{|lllc|} \hline
\multicolumn{4}{|c|}{\bf Reference intensity: $I_{0}$}
\\
\multicolumn{4}{|c|}{($M\geq 10^{10}$ \Msol, no subhalos)}
\\[0.1cm]
\multicolumn{1}{|c}{\em Physics properties} &
{\em Reference $I_{0}$} &
{\em Variations $I_{0,\,\mathrm{var}}$} &
$|I_0-I_{0,\,\mathrm{var}}|/I_0$
\\[0.2cm]
Halo mass function$^\dagger$ &
R16 \cite{2016MNRAS.462..893R} &
T08 \cite{2008ApJ...688..709T}, B16 \cite{2016MNRAS.456.2361B} &
$\lesssim \unit[40]{\%}$
\\
Density profile $\rho_{\mathrm{halo}}$  &
$\alpha_{\rm E}$  = 0.17 &
$\alpha_{\rm E}=0.15$, $\alpha_{\rm E}=0.22$, NFW &
$\lesssim\unit[20]{\%}$
\\
$c_{\Delta}(M_{\Delta})$ relation$^\ddagger$& C15 \cite{2015MNRAS.452.1217C}   &
L16 \cite{2016MNRAS.460.1214L}, C15-$\sigma_c$=$0.2$, \textcolor{gray}{(S14)} &
$\lesssim \unit[10]{\%}$
\\
Cosmology ($h,\,\Omega_i,\,P_k$)$^\S$  &
\planck{}--R16 \cite{2016MNRAS.462..893R}  &
WMAP7$\,$\cite{2011ApJS..192...18K},$\,$\textcolor{gray}{(WMAP-T08)}&
$\lesssim\unit[10]{\%}$
\\
Overdensity definition &
$\Delta_{\rm vir}$
(\ref{eq:bryannorman}) &
$\Delta_{\rm c}$\,(\ref{eq:deltacrit})\,or\,$\Delta_{\rm m}$\,(\ref{eq:deltam})=200 &
$\lesssim\unit[5]{\%}$
\\[0.1cm]
EBL model$^\star$ &
I13 \cite{2013ApJ...768..197I} &
F08 {\cite{2008A&A...487..837F}}, D11 \cite{2011MNRAS.410.2556D}, G12 \cite{2012MNRAS.422.3189G} & $\lesssim 5-\unit[40]{\%}$
\\[0.6cm]
\multicolumn{4}{|c|}{\bf Total CDM contribution: $I_{l}$ (extrapolation to low masses)}
\\
\multicolumn{4}{|c|}{($M\geq M_{\rm min}$, no subhalos)}
\\[0.1cm]
\multicolumn{1}{|c}{\em Field halo properties}    & \multicolumn{2}{l}{\em Values (default in \textbf{bold})}               & $I_{l}/I_0$ {\em ($\simeq$ 5)}   \\[0.2cm]
Slope of $\dd n/\dd M$, $\alphaM$           & \multicolumn{2}{l}{1.85, $\bm{1.9}$, 1.95}                  & $\sim4-14$    \\
Minimal mass $M_{\rm min}$                  & \multicolumn{2}{l}{$10^{-12}$, $\bm{10^{-6}}$, $\unit[10^{-3}]{\Msol}$} & $\sim4-8$ \\
Density profile $\rho_{\mathrm{halo}}$      & \multicolumn{2}{l}{$\alpha_{\rm E}  = 0.15,\, \bm{0.17},\, 0.22$, NFW, Ishiyama \cite{2014ApJ...788...27I}} & $\sim 4-8$ \\
$c_{\Delta}(M_{\Delta})$ relation$^\ddagger$& \multicolumn{2}{l}{
\textbf{C15} \cite{2015MNRAS.452.1217C}, L16 \cite{2016MNRAS.460.1214L}, \textcolor{gray}{(S14 \cite{2014MNRAS.442.2271S})}} & $\sim3-8$
\\[0.3cm]
\multicolumn{4}{|c|}{\bf \dots including boost from subhalos: $I_{\mathrm{b}}$}
\\[0.1cm]
\multicolumn{4}{|c|}{($m\geq m_{\rm min}$ with $m_{\rm min}\equiv M_{\rm min}$)}
\\[0.1cm]
\multicolumn{1}{|c}{\em (Sub-)halo properties}& \multicolumn{2}{l}{\em Values (default in \textbf{bold})}               & $I_{\mathrm{b}}/I_{l}$ {\em ($\simeq$ 1.5)}   \\[0.2cm]
Mass fraction \fsub{}                       & \multicolumn{2}{l}{10\,\%, $\bm{20\,}$\textbf{\%}, 40\,\%}                 &  $\sim1.2-2.2$ \\
Minimal mass $m_{\rm min}$                  & \multicolumn{2}{l}{$10^{-12}$, $\bm{10^{-6}}$, $\unit[10^{-3}]{\Msol}$} &  $\sim1.3-1.8$   \\
$c_{\Delta}(M_{\Delta})$ relation$^\ddagger$& \multicolumn{2}{l}{
\textbf{C15} \cite{2015MNRAS.452.1217C}, L16 \cite{2016MNRAS.460.1214L}, \textcolor{gray}{(S14 \cite{2014MNRAS.442.2271S})}} & $\sim1.3-1.7$ \\
Density profile $\rho_{\mathrm{subhalo}}$   & \multicolumn{2}{l}{$\alpha_{\rm E}  = 0.15,\, \bm{0.17},\, 0.22$, NFW, Ishiyama \cite{2014ApJ...788...27I}} & $\sim 1.3-1.7$ \\
Slope of $\dd P/\dd m$,  $\alpham$          & \multicolumn{2}{l}{1.85, $\bm{1.9}$, 1.95}                  &  $\sim1.4-1.7$   \\
$\dd P/\dd V$ profile                       & \multicolumn{2}{l}{\textbf{Aquarius} \cite{2008MNRAS.391.1685S}, Ph{\oe}nix \cite{2012MNRAS.425.2169G}, $\propto\rho_{\mathrm{host}}$} & $\sim 1.49-1.51$ \\\hline
\end{tabular}
 \\
 {\tiny
 $^\dagger$ T08 (Tinker et al., 2008), B16 (Bocquet et al., 2016), R16 (Rodr\'guez-Puebla et al., 2016)\\
 $^\ddagger$ S14 (S{\'a}nchez-Conde \& {Prada},  2014, \cite{2014MNRAS.442.2271S}), C15 (Correa et al., 2015), L16 (Ludlow et al., 2016)\\
 $^\S$ \planck{}--R16 (MultiDark--\planck{} simulations used in Rodr\'iguez-Puebla et al., 2016), WMAP--T08 (Cosmology used in T08, \cite{2008ApJ...688..709T})\\
 $^\star$ F08 (Franceschini et al., 2008), D11 (Dom\'inguez et al., 2011), Gilmore et al. (2012), and I13 (Inoue et al., 2013)\\
 }
 \caption{Parameters and options used for the calculations of the intensity, \cref{eq:mean_intensity}. The table is organised in three blocks, starting from the high-mass halo contribution $I_0$ ($M>\unit[10^{10}]{\Msol}$) discussed in \cref{subsec:benchmark_model}, the all-mass halo contribution $I_l$ where extrapolations of the parameters are used in the low-mass range (\cref{subsubsec:small_mass_divergence}), and finally $I_\mathrm{b}$ that accounts for substructures in the halos (\cref{subsubsec:substructure}). In more details, from top to bottom: (i) benchmark parameters, their alternatives, and induced uncertainties on the reference intensity $I_0$ (the models given in grey/parentheses are investigated but not included in the error budget -- see discussion and \cref{fig:DiffFluxRatioHMFuncertainties,fig:DiffFluxRatioHMFextrapolation,fig:DiffFluxRatioSubsUncertainties}); (ii) parameters extrapolated to low masses, range of their values, and impact on the ratio $I_l/I_0$ (see \cref{fig:DiffFluxRatioHMFextrapolation}); (iii) subhalo parameters, range of their values, and impact on the ratio $I_\mathrm{b}/I_l$ (see \cref{fig:DiffFluxRatioSubsUncertainties}). Note that for (ii) and (iii) the bold values correspond to the default configuration used when varying one parameter at a time. Also, the minimal mass, the concentration $c_{\Delta}(M_{\Delta})$, and halo profiles are always tied between field and subhalos.
}
\label{tab:benchmarkmodel}
\end{table}

\subsection{Contribution $I_0$ from high-mass halos ($M\geq\unit[10^{10}]{\Msol}$): a robust lower bound}
\label{subsec:benchmark_model}

In this first result section, we focus on the {\it safe} mass range $M\geq\unit[10^{10}]{\Msol}$, where constraints exist from both numerical simulations and observations. Also, we do not take into account substructures in these halos. The derived intensity will provide a lower limit on the DM contribution to the DGRB.

\begin{figure}[!ht]
\centering
\includegraphics[width=\textwidth]{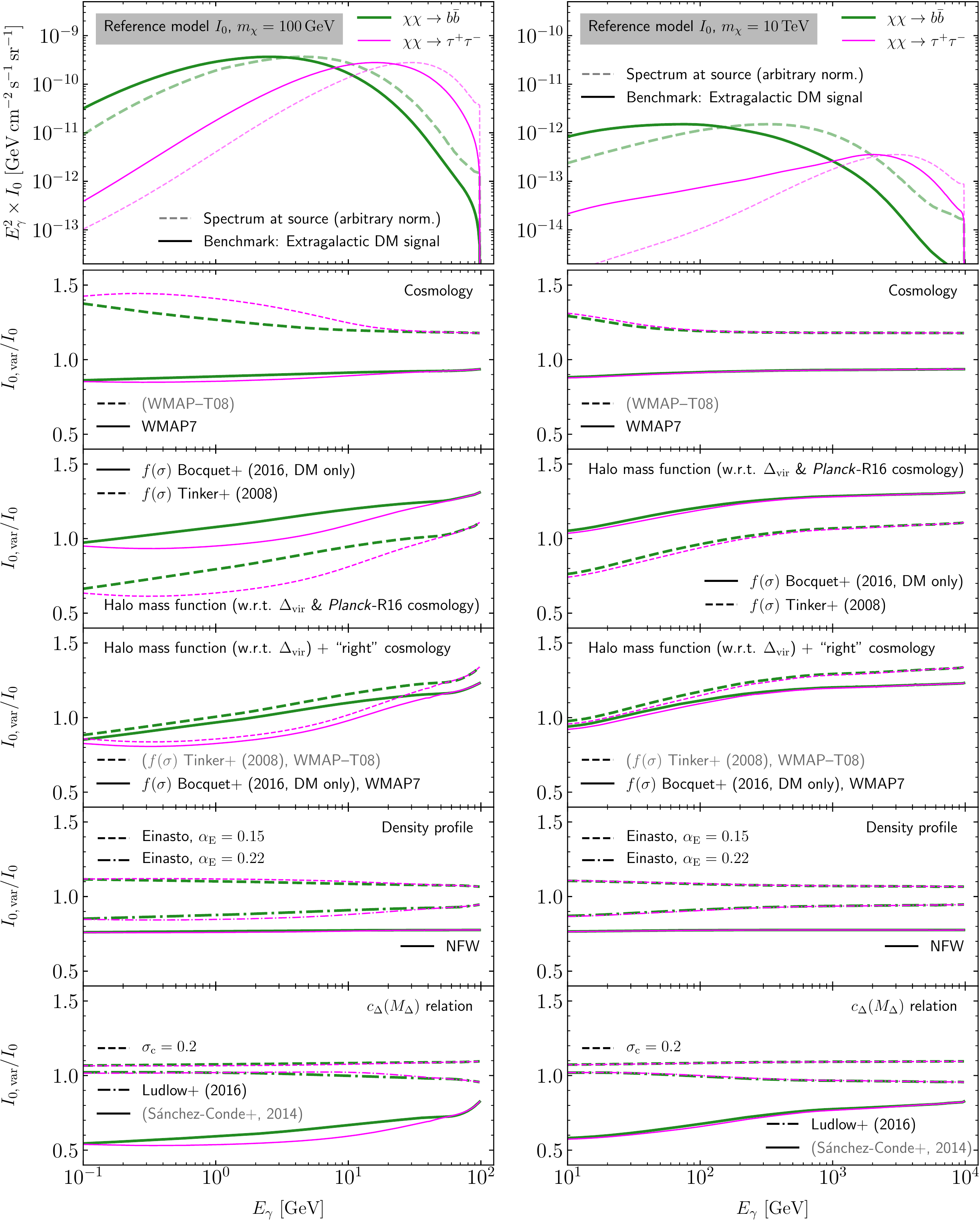}
\caption{{{\em Top row:} Reference intensity $I_0$ ($M\geq \unit[10^{10}]{\Msol}$, no subhalos) for $\sigmav=\unit[3\times 10^{-26}]{\frac{cm^3}{s}}$ and a light (\unit[100]{GeV}, {\em left}) or heavy (\unit[10]{TeV}, {\em right column}) WIMP. Pure $\chi\chi\rightarrow b\bar{b}$ (green) and $\chi\chi\rightarrow\tau^+\tau^-$ (magenta curves) annihilation channels are shown. Dashed lines in the top row display the source spectra without EBL absorption and cosmological redshift. {\em Lower rows:} Ratios $I_0$ to variations $I_{0,\rm var}$. Bracketed models  are excluded from the error budget. See \cref{tab:benchmarkmodel} and discussion in \cref{subsec:benchmark_model}.
}
\label{fig:DiffFluxRatioHMFuncertainties}
}
\end{figure}

The top third of \cref{tab:benchmarkmodel} (denoted {\it reference intensity}) and \cref{fig:DiffFluxRatioHMFuncertainties} summarise the results that will be discussed in this section. In this figure, we present the exotic extragalactic intensity for DM candidates of 100\;GeV (left column) and 10\;TeV (right column). While a 100\;GeV particle corresponds to the canonical mass scale of a generic weakly interacting massive particle (WIMP), $\mchi=\unit[10]{TeV}$ marks the regime of the largest expected WIMP masses \cite{1674-1137-40-10-100001}. Additionally, these two candidates allow us to explore regimes of weak and strong EBL absorption of remote \grs{}. For illustrative purpose, we also always display results for the $b\bar b$ (green) and the harder $\tau^+ \tau^-$ (magenta) annihilation channels. The top row in \cref{fig:DiffFluxRatioHMFuncertainties} corresponds to the intensity from the reference model, while lower panels show the deviation obtained by changing the ingredients of the default configuration; this is discussed in the following paragraphs.

\paragraph{Cosmology.}
The extragalactic exotic \gr{} intensity given by \eq{eq:mean_intensity} depends on cosmology through today's mean DM density, the Hubble expansion rate and the halo mass function. The latter depends not only on the cosmology but also on the specific parameters of the multiplicity function that are fitted to results of numerical simulations. We first evaluate the impact of the cosmological parameters alone, using the sets of parameters given in \cref{table:cosmo_params}. The latter correspond to the underlying cosmologies of several simulations from which the mass functions discussed below have been derived. We use the \planck{}--R16 cosmology as our reference, and study the impact on the intensity when switching to the WMAP7 or WMAP--T08 cosmologies. The results are displayed in the second row of \cref{fig:DiffFluxRatioHMFuncertainties}.
\begin{table}
\centering
\begin{tabular}{lccccccc} \hline\hline
                     & $h $& $\Omega_{\rm m,0}$  &  $\Omega_{\rm b,0}$  &  $\Omega_{\Lambda,0}$ & $\sigma_8$ & $n_s$ \\
 \hline
\planck{}--R16  & 0.678 & 0.307 & 0.048 & 0.693 &0.829 & 0.96 \\
WMAP7       & 0.704 & 0.272 & 0.0456& 0.728 & 0.809& 0.963 \\
WMAP--T08  & 0.7     & 0.3     & 0.04   & 0.7     & 0.9    & 1  \\
\hline
\end{tabular}
\caption{Cosmological parameter sets considered in this study with $h=H_0/100\times\unit{Mpc\;s\;km^{-1}}$, $\varrho_{i}(z=0)=\Omega_{i,0}\times\varrho_\mathrm{c}(0)$,  $\sigma_8 = \sigma(R=\unit[8\,h^{-1}]{Mpc})$, and $n_\mathrm{s}$ the spectral index of the primordial power spectrum. \planck{}--R16 corresponds to the \planck{} cosmology as implemented in the MultiDark--\planck{} and Bolshoi--\planck{} simulations \cite{2016MNRAS.457.4340K} used by Rodr\'iguez-Puebla et al. (2016, R16 \cite{2016MNRAS.462..893R}). The WMAP7 cosmology was implemented in the Magneticum simulations used by Bocquet et al. (2016, B16 \cite{2016MNRAS.456.2361B}). Finally, WMAP--T08 corresponds to one of the WMAP1--3 cosmologies implemented in the simulations used by Tinker et al. (2008, T08 \cite{2008ApJ...688..709T}).\label{table:cosmo_params}}
\end{table}

When using the WMAP--T08 cosmology, the effect can reach $\sim 50\%$ for the 100\;GeV DM candidate; this is shown for illustrative purposes only, as this outdated set of parameters is ruled out by more recent estimates. Apart from this case, switching between \planck{} and WMAP7 cosmologies, the impact on the intensity remains a rather marginal $\sim 10\%$ effect, and we only propagate this uncertainty to our total error budget. Note that the effect is even slightly smaller for the \unit[10]{TeV} DM candidate (right) than for the 100\;GeV candidate (left). For TeV dark matter, the ratio to the reference intensity is also rather independent of the annihilation channel; at these high energies the EBL absorption is such that the \grs{} have a local origin only, i.e. their spectra are not redshifted and the ratio to the reference intensity is therefore the same for both channels. This behaviour will be present in all the cases we explore below.

\paragraph{Halo mass function/multiplicity function.} We now turn to the choice of the parametrisation of the multiplicity function $f(\sigma,z)$ entering the calculation of the mass function $\dd n/\dd M$ according to \cref{eq:halomassfunction}. In the last fifteen years, the parametrisation of $f(\sigma,z)$ has evolved, following the improvements of cosmological simulations \cite{2012PDU.....1...50K}. As mentioned above, for this work we select the recent Rodr\'iguez-Puebla et al. parametrisation (R16, \cite{2016MNRAS.462..893R}) as our reference mass function. In order to bracket the modelling uncertainties, we also consider the widely-used Tinker et al. (T08, \cite{2008ApJ...688..709T}) and the Bocquet et al. (B16, \cite{2016MNRAS.456.2361B}) DM-only parametrisations.\footnote{We validate our implementation of the cosmology and mass functions in the \clumpy{} code  from a successful comparison to the original publications, i.e. using the same underlying cosmology as the one the mass function was derived from (see \cref{table:cosmo_params} and \cref{fig:HMF_comparison_threeInRow} in \cref{sec:intensity_multiplier_bricks}).} In B16, the authors also provide results based on hydrodynamical simulations, including baryon feedback on structure formation. Using these results yield only a 10\% difference on the exotic extragalactic intensity compared to the DM-only parametrisation, so we only consider the latter below.

The third and fourth rows in figure~\ref{fig:DiffFluxRatioHMFuncertainties} show the impact of the choice of multiplicity function when (i) the \planck{} cosmology is used whatever the chosen multiplicity function or when (ii) a given  $f(\sigma,z)$ is combined with the ``right'' cosmology, i.e. the one of the simulations used to determine its parameters. In both cases, there is at most a $\sim 40 \%$ difference with respect to the reference intensity.

\paragraph{Overdensity definition.}
The relation between a halo size and its mass is given by \cref{eq:rdeltamdelta} and depends on the overdensity quantity $\Delta(z)$. Various definitions are found in the literature, namely
\begin{align}
\Delta(z) & = {\rm const.} =: \Delta_{\rm c}\label{eq:deltacrit},\\
\Delta(z) &= {{\rm const.} \times \Omega_{\rm m}(z)} =: {\Delta_{\rm m} \times \Omega_{\rm m}(z)}\label{eq:deltam},\\
\Delta(z) &= 18\pi^2 + 82\,[\Omega_{\rm m}(z) - 1)]- 39\,[\Omega_{\rm m}(z) -
1]^2=: \Delta_{\mathrm{vir}}\quad(\text{\cite{1998ApJ...495...80B}, for a flat
Universe})\label{eq:bryannorman}\,.
\end{align}
Halo mass functions are generally given for several values of $\Delta(z)$, common choices being $\Delta_{\rm {c,m}}=200$. Although not shown in \cref{fig:DiffFluxRatioHMFuncertainties}, we also explore how the results are affected by this choice. The methodology to convert the halo mass between different $\Delta(z)$ is described in \cref{sec:mdeltaconversion}. Using $\Delta_{\rm vir}$ as the reference, we find a very marginal effect  of $\lesssim 5\%$ when switching to $\Delta_{\rm c}=200$ or $\Delta_{\rm m}=200$.

\paragraph{Halo density profile.} The halo profile enters into the intrinsic luminosity term \cref{eq:luminosity}. Two standard, spherically-symmetric\footnote{DM halos are not spherical, the more massive halos being more triaxial \cite[e.g.,][]{2015MNRAS.449.3171B,2017MNRAS.466..181D}. However, as we average on large numbers of halos and orientations, we do not expect triaxiality to impact the results.} parametrisations of DM profiles are the NFW \cite{1996ApJ...462..563N} and Einasto \cite{1989A&A...223...89E} profiles,
  \begin{align}
    \rho^{\rm NFW}(r)&=\frac{4\,\rho_{-2}}{(r/r_{-2})(1+r/r_{-2})^2},
         \label{eq:NFW}\\[0.2cm]
     \rho^{\rm Einasto}(r)&=\rho_{-2} \exp\left(-\frac{2}{\alpha_{\rm E}}\left[\left(\frac{r}{r_{-2}}\right)^{\alpha_{\rm E}} -1\right]\right),
     \label{eq:Einasto}
  \end{align}
with $r_{-2}$ the radius for which the logarithmic slope equals $-2$, and $\rho_{-2}=\rho(r_{-2})$ the normalisation. The NFW profile has an inner slope of $-1$, whereas the Einasto profile logarithmically tends to a flat profile, with the sharpness of the inner profile controlled by $\alpha_{\rm E}$; the smaller $\alpha_{\rm E}$, the steeper the slope.

The Einasto parametrisation has been found to better fit DM halos than the NFW profile, in both DM-only \cite{2008MNRAS.391.1685S,2009MNRAS.398L..21S,2014MNRAS.441.3359D} and hydrodynamical simulations \cite{2017MNRAS.465.3361H}, and at various scales. A large suite of Milky-Way size simulated halos ($M\sim \unit[10^{12}]{\Msol}$) obtained $\alpha_{\rm E}\approx 0.17\pm0.02$ \cite{2016ApJ...818...10G}, in agreement with the results of \cite{2008MNRAS.391.1685S}. However, this slope is not universal, and dedicated DM-only simulations found that $\alpha_{\rm E}$ increases with both the mass and redshift \cite{2008MNRAS.387..536G,2014MNRAS.441.3359D,2017MNRAS.465L..84L}. In particular between $M=\unit[10^{11}]{\Msol}$ and $M=\unit[10^{16}]{\Msol}$, \cite{2014MNRAS.441.3359D} finds that $\alpha_{\rm E}$ increases from 0.16 to 0.22. At these high masses, the hydrodynamical feedback from active galactic nuclei in the halo centers can affect the profile \cite{2017MNRAS.472.2153P}. From the observational point of view, lensing constraints have found $0.17<\alpha_{\rm E}<0.21$ for halos with $M\sim \unit[10^{15}]{\Msol}$ \cite{2017ApJ...836..231U}, whereas X-ray data analyses found somewhat a larger range of values with $0.14<\alpha<0.26$ \cite{2011ApJ...736...52H} or $\alpha\approx 0.29$ \cite{2016MNRAS.462..681M}.

For our reference calculation of $I_0$, we assume an Einasto profile of index $\alpha_{\rm E}=0.17$. This value is more representative of Milky-way like halos and small groups than galaxy clusters, but the most massive ones are not numerous and are subordinate in the \gr{} signal (see the sharp decrease of the mass function above $\unit[10^{14}]{\Msol}$ at $z=0$ in \cref{fig:HMF_comparison_extrapolation}). We do not include any mass and redshift dependence in $\alpha_{\rm E}$, but $\alpha_{\rm E}$ is varied from 0.15 to 0.22 to encompass the possible values obtained in the simulations and the data. The impact on the \gr{} intensity is shown in the fifth row of \cref{fig:DiffFluxRatioHMFuncertainties}, where a $10-15\%$ effect is observed. We also show the comparison with a NFW profile, which is $\sim20\%$ below the  Einasto reference profile, but close to $\alpha_{\rm E}=0.22$. This is both in agreement with the fact that Einasto profiles with this index are close to NFW profiles, and the fact that despite they asymptotically flatten, they give a larger \gr{} signal; this is because they produce larger densities than NFW profiles of same mass and mass concentration in regions which dominate the signal from annihilation.

\paragraph{Mass-concentration-redshift parametrisation $c(M,z)$.}
Given a halo mass, a profile (Einasto or NFW), and a $\Delta$ definition, the mass-concentration relation \cref{eq:C-M} fully determines the structural parameters $r_{-2}$ and $\rho_{-2}$ in \cref{eq:NFW,eq:Einasto}. This relation reflects the mass dependence of halo formation times, with less concentrated halos at earlier times. Early studies proposed a redshift dependence of $c(M,z)\propto(1+z)^{-1}$ \cite{2001MNRAS.321..559B,2001ApJ...554..114E}, but a milder dependence was obtained in subsequent calculations for high-mass halos \cite{2003ApJ...597L...9Z}. Several empirical models have been proposed since, taking advantage of the connections between halo mass profiles and the main progenitor mass accretion history.

We rely here on three models \cite{2014MNRAS.442.2271S,2015MNRAS.452.1217C,2016MNRAS.460.1214L}, namely S14 (S{\'a}nchez-Conde \& {Prada},  2014), C15 (Correa et al., 2015), and L16 (Ludlow et al., 2016). The domain of validity of these models encompasses a large mass range, providing a consistent picture when extrapolating the mass function down to the lower masses (see \cref{subsec:lowmass_contrib}). Moreover, the recent works by C15 and L16 include a dedicated redshift dependence study of the mass-concentration, which makes them appealing for the calculation of the {\gr} emission from far-away DM structures. While the S14 model was obtained for $z=0$ and is extended to higher redshifts by a $(1+z)^{-1}$ scaling, C15 and L16 {\rm found} a  milder evolution: also, S14 shows an upturn for the highest masses contrarily to the other two (see the comparison in \cref{fig:Concentrations_1+z} in \cref{sec:intensity_multiplier_bricks}). Whether this upturn is real or a selection bias in numerical simulations is still under discussion \cite{2016MNRAS.457.4340K,2016MNRAS.460.1214L}.\footnote{While such an upturn is generally expected to provide a minor contribution to the \gr{} signal  as the number of very massive halos is strongly suppressed at high masses, it involves using the corresponding concentration relation beyond its fitting range.} At the galaxy cluster scale, the $c(M,z)$ from simulations has been found in agreement with X-ray and weak-lensing constraints, $c_{\rm vir}\sim 3-6$ \cite{2015ApJ...806....4M}, with a log-normal distribution of intrinsic scatter $\sigma_c\sim 0.12-0.22$ \cite{2015ApJ...806....4M,2016MNRAS.462..681M,2016A&A...590A.126A,2017arXiv170807349B} also consistent with results from numerical simulations \cite{2004A&A...416..853D,2014MNRAS.442.2271S}. Observations up to $z\sim1.2$ have shown no obvious redshift evolution \cite{2016A&A...590A.126A}, but more data are required to be more conclusive.

The C15 concentration is used as reference. The last panel of \cref{fig:DiffFluxRatioHMFuncertainties} shows the ratio of  the \gr{} intensity to this reference for different concentration choices, using S14, L16, or applying a log-normal scatter to the C15 model (C15-$\sigma_c$). The options L16 and C15-$\sigma_c$ give results within 10\% of the reference. The \gr{} intensity from S14 is markedly below (up to 50\% for the lowest energies), however, this difference is ascribed to the $(1+z)^{-1}$ evolution. Therefore, we do not consider the S14 model for our overall error budget and suggest a lower uncertainty on the intensity $I_0$ of not more than 10\%.

\paragraph{EBL absorption.} Intergalactic low-energy radiation fields absorb high-energy \grs{} via production of $e^+ e^-$ pairs. The intensity of the extragalactic infrared- and optical background light can either directly be estimated by photometric measurements, integration over deep-exposure galaxy counts, or indirectly by VHE observations of distant blazars (see \cite{2016RSOS....350555C} for a recent review). While different methods and measurements are hampered by different uncertainties and are able to give lower (galaxy counts) or upper (indirect \gr{} measurements) limits on the EBL density, they mostly agree in determining the spectral EBL energy density at low redshifts, $z<2$; the largest uncertainties are in the far-infrared, where zodiacal light is dominant and  astronomical observation intricate \cite{2001ARA&A..39..249H}. They significantly differ at redshifts $z>2$, caused by different extrapolations and evolution of the measurements into the past.  For this work, we apply as default the model from \cite{2013ApJ...768..197I}, I13 (Inoue et al., 2013), where they attempt for the first time to consistently calculate, using semi-analytical models of hierarchical structure formation of dark and baryonic matter, the EBL density at redshifts back to the epoch of reionization. This allows us to integrate \cref{eq:mean_intensity} up to $z_{\rm max}=10$. As their modelling predicts a factor $\sim 2$ lower \gr{} attenuation for $E_\mathrm{e}\gtrsim\unit[400]{GeV}$ \grs{} compared to previous calculations, we compare their estimation of the cosmic \gr{} opacity with the classical, more data-driven models by \cite{2011MNRAS.410.2556D} (D11, Dom\'inguez et al., 2011) based on observed galaxy populations up $z=1$ (and their evolution into the past to $z=2$), the backward-evolution model from \cite{2008A&A...487..837F} providing EBL attenuations up to $z=4$ (F08, Franceschini et al., 2008), and the semi-analytic  forward-evolution model by \cite{2012MNRAS.422.3189G} up to $z=6$ (G12, Gilmore et al., 2012). At higher redshifts, we extrapolate their models by a power-law, however, this only affects the lowest energies $E_{\rm \gamma}\lesssim \unit[20]{GeV}$ ($E_{\rm \gamma}\lesssim \unit[10]{GeV}$), for which the Universe is transparent enough to \grs{} emitted at $z\gtrsim 2$ ($z\gtrsim4$).

\begin{figure}[t]
\begin{flushright}
\includegraphics[width=0.985\textwidth]{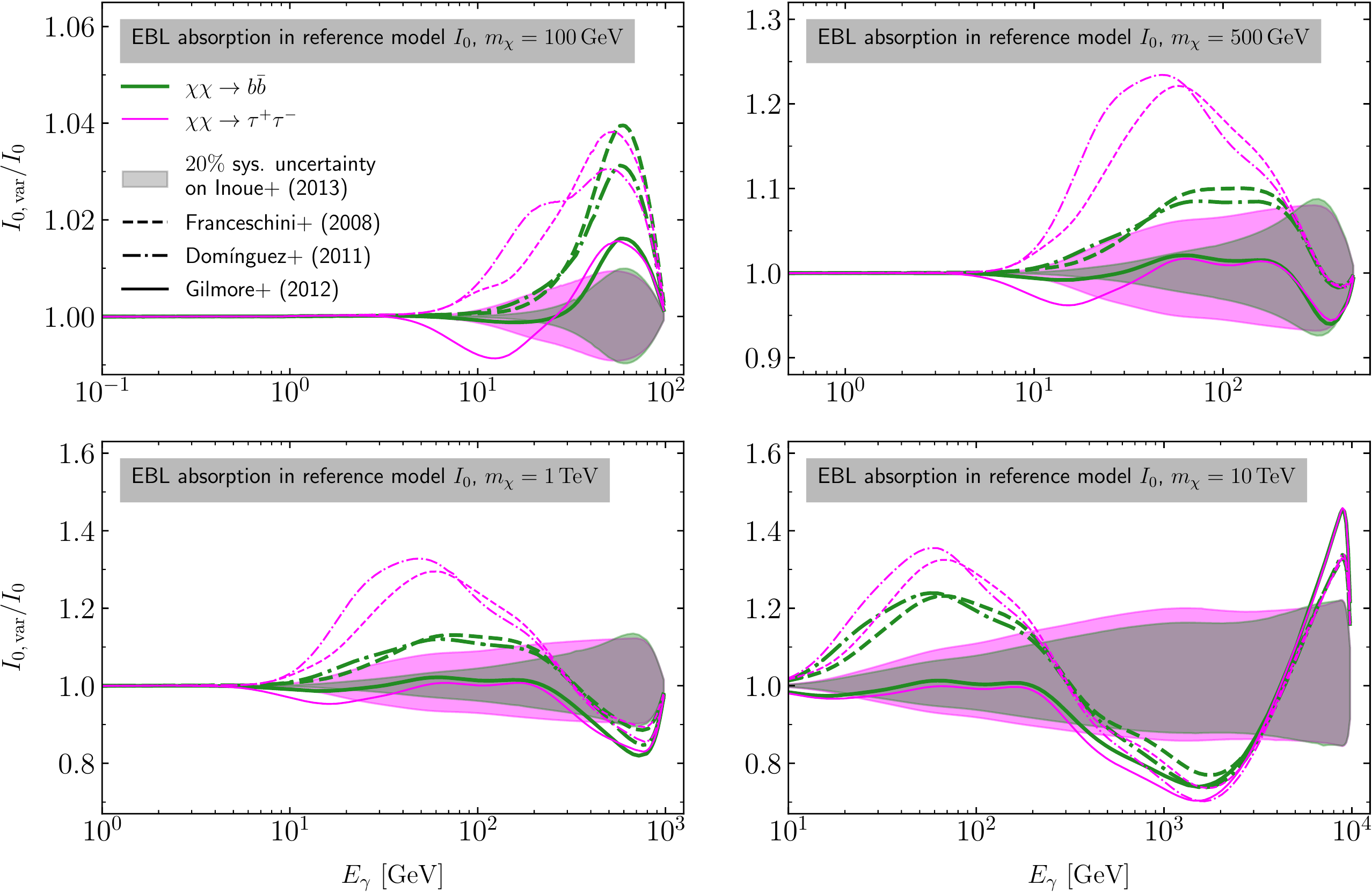}
\caption{{
Uncertainty of the EBL extinction factor $\tau(E_{\gamma},z)$ for four different WIMP masses between \unit[100]{GeV} and \unit[10]{TeV} and pure $\chi\chi\rightarrow b\bar{b}$ (green) and $\chi\chi\rightarrow \tau^+\tau^-$ (magenta) annihilation channels. The shaded areas indicate a $20\%$
systematic uncertainty on  $\tau(E_{\gamma},z)$ in the reference model \cite{2013ApJ...768..197I}.
}
\label{fig:DiffFluxRatioEBLuncertainties}}
\end{flushright}
\end{figure}
In \cref{fig:DiffFluxRatioEBLuncertainties}, we compare results from these four models for four different DM masses. As has been stated by the authors themselves, at low energies, the model I13 agrees with G12 and predicts a larger attenuation than F08 and D11, while it significantly differs from all compared models at energies above $E_\gamma\gtrsim\unit[200]{GeV}$. For a better assessment of the discrepancy, we also show the range of a $20\,\%$ systematic uncertainty on the attenuation factor $\tau$ in the I13 model. Overall, we find that the uncertainty from different models of the EBL increases with the \gr{} energy, from a marginal discrepancy at $E_\gamma\lesssim\unit[50]{GeV}$  up to a $40\,\%$ uncertainty at the highest energies, where the absorption is strongest. In general, the uncertainty is larger for a hard annihilation channel like the $\chi\chi\rightarrow \tau^+\tau^-$ case, with a larger relative amount of photons emitted at the high-energy end of the spectrum.

\subsection{Including low-mass halos and subhalos}
\label{subsec:lowmass_contrib}
\subsubsection{Extrapolation of the mass function}
\label{subsubsec:small_mass_divergence}

Cosmological simulations within the $\Lambda$CDM paradigm only determine halo structure down to halo masses of $M\gtrsim \unit[10^{10}]{\Msol}$. The number density of DM halos below the resolution limit is largely unknown, with however a major impact on the overall  \gr{} intensity from DM annihilations. Overall, three quantities govern the  \gr{} emission from small-scale DM structures: the number density of halos below the resolution limit, their minimal mass, and a possible higher mass concentration of lighter halos, which may further enhance the DM annihilation process.

The first row of \cref{fig:DiffFluxRatioHMFextrapolation} shows, for the reference configuration (see \cref{tab:benchmarkmodel}), the intensity accounting for the extrapolated low-mass field halos, $I_l$, compared to the reference calculation $I_0$. DM halos below $10^{10}$~\Msol{} dominate the intensity with $I_l/I_0\sim 5$; the contribution of various halo masses and redshifts are explicit in the right panel of \cref{fig:Fluxmultiplier_luminosities} in \cref{sec:intensity_multiplier_bricks}.

\paragraph{Power-law mass function extrapolation.}

For scale-invariant primordial perturbations, the matter power spectrum in a collisionless CDM paradigm approaches $P_{\rm lin}(k)\propto k^{-3}\times \ln^2(k)$ for $k\rightarrow \infty$ \cite{2014MNRAS.441.1861S}, and Press-Schechter theory predicts a corresponding power-law scaling of the mass function,
\beq
\frac{\dd n}{\dd M} \propto M^{-\alphaM}\quad\text{for $M\rightarrow 0$, }
\label{eq:pl_hmf_Extrapolation}
\eeq
with $\alphaM=2$ \cite{2008gady.book.....B}. We therefore adopt \cref{eq:pl_hmf_Extrapolation} to smoothly extrapolate the mass function below the minimal mass of our reference model, i.e. for $M<\unit[10^{10}]{\Msol}$. However, assuming $\alphaM=2$ and the mass functions listed in \cref{tab:benchmarkmodel}, the comoving density of mass contained in halos, $\varrho_{\rm halos}(z)=\int_{M_{\rm min}}^\infty M\,\dd n(z)/\dd M \,\dd M$, exceeds the mean density $\overline{\varrho}_{\rm m, 0}$ with $M_{\rm min}$ far above the natural mass cut-off scale (discussed in the next paragraph).\footnote{At $z\lesssim 0.5$, we get $\varrho_{\rm halos} = \overline{\varrho}_{\rm m, 0}$ for $\dd n/\dd M \propto M^{-2}$ at $M_{\rm min}\gtrsim \unit[1]{\Msol}$, this number depending on the cosmology and choice of $f(\sigma)$. At higher redshifts, before the dominance of $\Omega_\Lambda$, $M_{\rm min}(\varrho_{\rm halos}(z)\equiv\overline{\varrho}_{\rm m, 0} )$ quickly decreases.} On the other hand, we obtain $\varrho_{\rm halos}(M_{\rm min}= \unit[10^{-12}]{\Msol}) \leq \overline{\varrho}_{\rm m, 0}$  with $\alphaM=1.95$ for all considered cosmologies and mass functions. By this argument we choose $\alphaM=1.95$ as an upper bound to consistently explore different minimal particle-physics motivated cut-off scales.
\begin{figure}[t]
\centering
\includegraphics[width=0.79\textwidth]{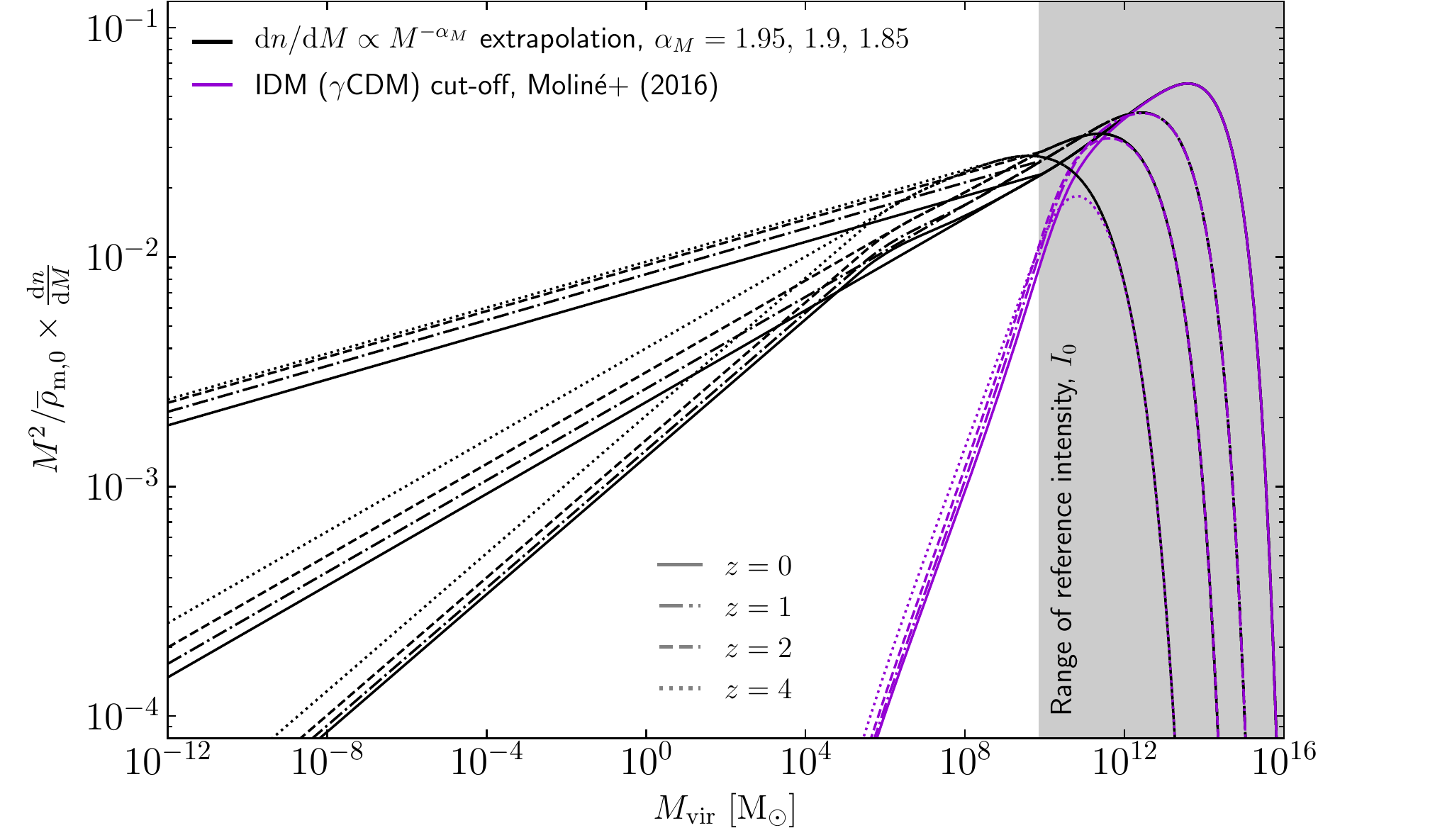}
\caption{
Halo mass function from R16 \cite{2016MNRAS.462..893R} with various assumptions of the low-mass scales. For the power-law extrapolations, we retain the logarithmic slope towards lower masses below $M_{\rm vir} = \unit[10^{10}]{\Msol}$ from where the original mass function slope adopts the value of \alphaM{}. The IDM model (purple curves) is defined by the mass function cut-off according to Eq.~(11) in \cite{2016JCAP...08..069M}.}
\label{fig:HMF_comparison_extrapolation}
\end{figure}
As a lower bound on \alphaM{}, we select $\alphaM=1.85$, which is the typical low-mass asymptotic slope in simulations \cite{2008ApJ...688..709T}. As can be seen in \cref{fig:HMF_comparison_extrapolation},  this slight decrease of the exponent significantly reduces the small-scale halo occupation and approaches the regime of alternative DM scenarios without any appreciable small-scale structures, which is commented on in \cref{subsec:SIDM}.

\begin{figure}[t]
\centering
\includegraphics[width=\textwidth]{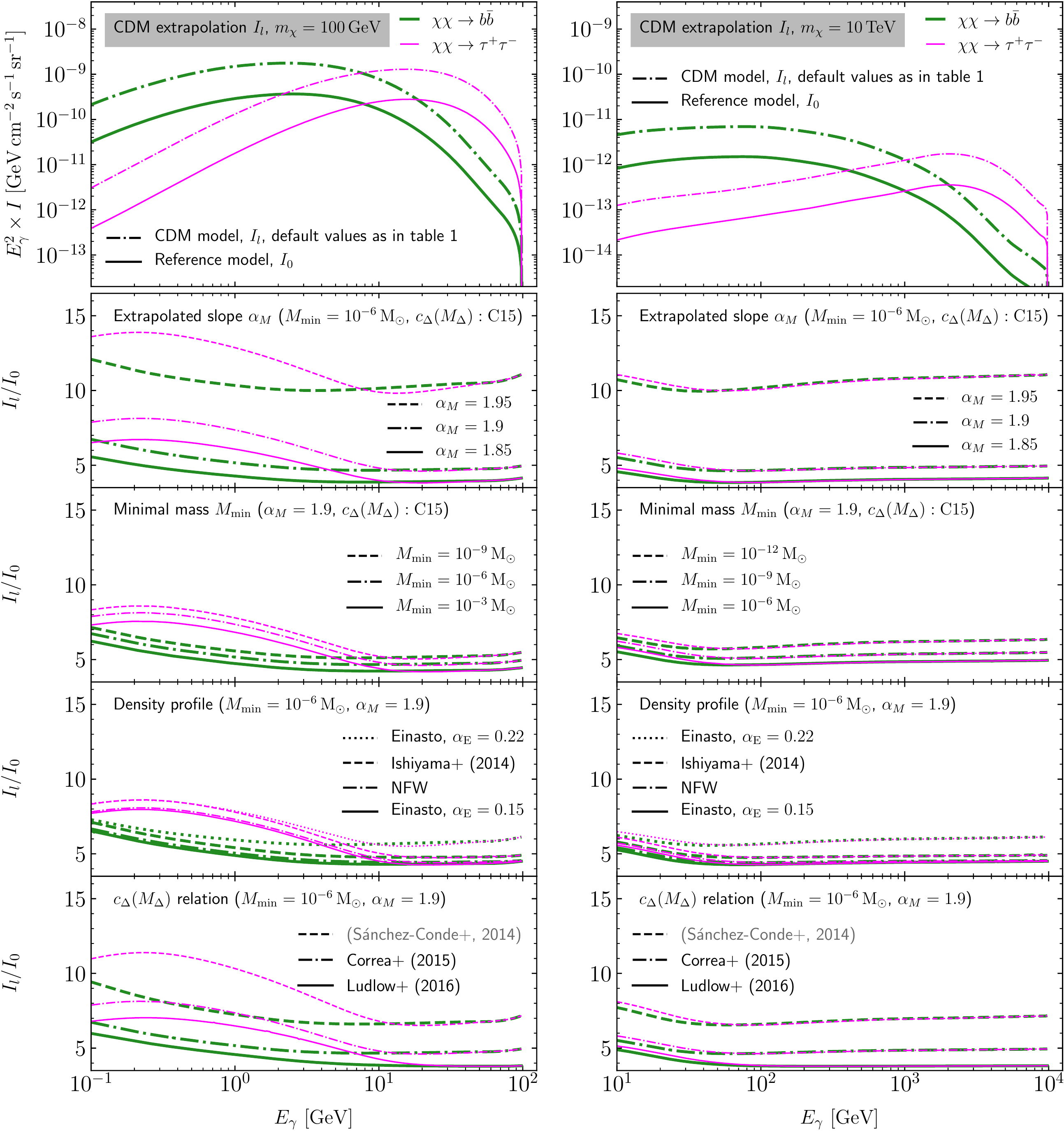}
\caption{{\em Top row:} Intensity $I_l$ (all halos, no subhalos) for the same WIMP configurations as in \cref{fig:DiffFluxRatioHMFuncertainties}, where we also report $I_0$ for comparison purpose. {\em Remaining rows:} Ratios of the $I_l/I_0$ varying field halo properties (\cref{tab:benchmarkmodel}). The bracketed model is excluded from the error budget. See discussion in \cref{subsec:lowmass_contrib}.}
\label{fig:DiffFluxRatioHMFextrapolation}
\end{figure}
The second row of \cref{fig:DiffFluxRatioHMFextrapolation} shows the drastic impact of \alphaM{}. A minimal mass of $M_{\rm min}= \unit[10^{-6}]{\Msol}$ and $\alphaM{}=1.85$ quadruples the signal from our reference model (solid curves in the lower panels of \cref{fig:DiffFluxRatioHMFextrapolation}), $I_l/I_0\sim 4$. The signal is slightly more enhanced at energies $\lesssim \unit[50]{GeV}$, where absorption by the EBL is  insignificant and signal from higher redshift halos may contribute to the signal. Steepening \alphaM{} to our default value of 1.9 further increases the signal by additional $20\,\%$ (dot-dashed curves \cref{fig:DiffFluxRatioHMFextrapolation}). With $\alphaM=1.95$, the increase of the signal is much larger as it is doubled compared to $\alphaM=1.9$ (dashed curves, $I_l/I_0\sim 10$). Note, however, that combined with a small $M_{\rm min}$ and high low-mass halo concentrations $c_\Delta$, $\alphaM >1.9$ may further enhance the \gr{} emission; for instance, for $\alpha_M=1.95$, a cut-off mass $10^{-12}$ \Msol{} would give $I_l/I_0\sim 30$.

\paragraph{Minimal halo mass, $M_{\rm min}$.}
In the early Universe, the kinetic decoupling of WIMPs \cite{2001PhRvD..64h3507H,2003PhRvD..68j3003B,2004MNRAS.353L..23G,2005JCAP...08..003G,2007JCAP...04..016B,2009NJPh...11j5027B,2014PhyU...57....1B} and to a lesser extent acoustic oscillations \cite{2005PhRvD..71j3520L,2006PhRvD..74f3509B} sets a small-scale cut-off on the mass of the smallest protohalos that can form. Numerical simulations have confirmed that such subhalos might survive until today \cite{2005Natur.433..389D}. A consistent calculation of the minimal mass associated with specific WIMP candidates was discussed in \cite{2006PhRvL..97c1301P,2009NJPh...11j5027B}, finding a range $[10^{-12}-10^{-4}]$ \Msol{}. The third row in \cref{fig:DiffFluxRatioHMFextrapolation} shows the impact of changing this minimal mass. Following \cite{2009NJPh...11j5027B}, we take a cut-off mass in $[10^{-9}-10^{-3}]$ \Msol{} for $m_\chi=100$~GeV (left panels) and $[10^{-12}-10^{-6}]$ \Msol{} for $m_\chi=10$~TeV (right panel). For $\alpham=1.9$, going down to smaller cut-off masses does only slightly further increase $I_l/I_0$, but would have a more drastic effect associated with a larger mass slope $\alpham\gtrsim1.95$, as has been discussed in the previous paragraph (see \cref{fig:Fluxmultiplier_luminosities} in \cref{sec:intensity_multiplier_bricks} for the flux multiplier per halo mass).

\paragraph{Inner profile of micro-halos.}
Whereas the slope of dark matter halos seems robustly determined at late stages of their evolution (see \cref{subsec:benchmark_model}), recent studies have shown that halos close to the free streaming scale could be cuspier than the NFW profile, with slopes as steep as $-1.5$ \cite{2010ApJ...723L.195I,2013JCAP...04..009A,2014ApJ...788...27I,2016MNRAS.461.3385O,2017MNRAS.471.4687A,2017arXiv170707693O}. We calculate $I_l$ for extreme Einasto slopes (0.15 and 0.22), NFW, and the micro-halo model of Ishiyama (2014), see Eq.~(2) in \cite{2014ApJ...788...27I}. In the next-to-last row of \Cref{fig:DiffFluxRatioHMFextrapolation}, we show $I_l/I_0$ in which the same profile model is used in $I_0$ and $I_l$ for consistency. The Ishiyama case which corresponds to NFW profiles for high-mass halos and steeper profiles for low-mass halos gives a $\sim 15\%$ increase w.r.t. the NFW case. The two Einasto cases\footnote{The steeper $\alpha_E=0.15$ profiles give a smaller $I_l/I_0$ than $\alpha_E=0.22$ profiles, contrarily to the ordering seen in \cref{fig:DiffFluxComparison}. This is because both $I_l$ and $I_0$ are changed here.} encompass all the other configurations with $I_l/I_0\sim 5-9$.

\paragraph{Extrapolated mass-concentration-redshift parametrisation.}
The models of \cite{2001MNRAS.321..559B,2001ApJ...554..114E}, extrapolating $c(M,z)$ down to the lowest masses, have been widely used alternatives to estimate annihilation signals for charged and neutral particles \cite[e.g.][]{2002PhRvD..66l3502U,2008A&A...479..427L,2012MNRAS.425..477N}. In recent years, refined semi-analytical models were able to reproduce a wide range of simulations in different cosmologies \cite{2015MNRAS.452.1217C,2016MNRAS.460.1214L}. We use the ready-to-use parametrisations App.~B1 of \cite{2015MNRAS.452.1217C} and App.~C of \cite{2017MNRAS.465L..84L} appropriate for a \planck{} cosmology. We refer the reader to these papers and references therein for systematic comparisons with previous $c(M,z)$ relations. In the last row of \Cref{fig:DiffFluxRatioHMFextrapolation}, we use C15 \cite{2015MNRAS.452.1217C} as default and compare to L16 \cite{2017MNRAS.465L..84L}, but also to S14 \cite{2014MNRAS.442.2271S} to study the impact of having a different redshift dependence. We find that L16 is $\sim20\%$ below C16, while S14 is $\sim40\%$ above. There is no simple explanation of these differences as they result from the interplay of crossing concentrations at different masses and redshifts for the various models (see \cref{fig:Concentrations_1+z} in \cref{sec:intensity_multiplier_bricks}).

\subsubsection{Boost from DM halo substructures}
\label{subsubsec:substructure}

In the hierarchical structure formation, host halos contain a population of subhalos with sub-subhalos, etc. down to the smallest scale. The annihilation signal from each individual halo is boosted from these populations. The boost increases with the mass \cite{2014MNRAS.442.2271S}, with only mild boosts $\lesssim 1.5$ for small halos and boost values up to 10 \cite{2012MNRAS.425..477N} or 100 \cite{2012MNRAS.425.2169G} for galaxy clusters. In \clumpy{}, the boost can be calculated for any distribution of subhalos down to any level of sub-substructures. We only consider the first level in the results below, as next levels only marginally contribute to the overall signal (e.g., Figure~1 of \cite{2016CoPhC.200..336B}). The boost depends on the subhalo properties, i.e. their number, their mass and spatial distributions, and their DM profiles. Dedicated numerical simulations of the subhalo population have been performed for Milky-Way like galaxies \cite{2007ApJ...657..262D,2008MNRAS.391.1685S,2016MNRAS.457.3492H,2017MNRAS.471.1709G} and galaxy clusters \cite{2012MNRAS.425.2169G,2016MNRAS.462..893R,2017MNRAS.469.4157C}, but they cannot explore subhalo masses below $M\sim\unit[10^{5}]{\Msol}$. Hence, for the boost calculation, low-mass subhalo properties must also be extrapolated down to the minimum subhalo mass used in the calculation. We refer the reader to \cite{2016JCAP...09..047H} for a thorough discussion of the subhalo properties and likely range of their parameters in the context of dark clumps detection in the Milky-Way at $z=0$. Here, we need in addition to consider the redshift dependence and host halo mass dependence.

\begin{figure}[t]
\centering
\includegraphics[width=\textwidth]{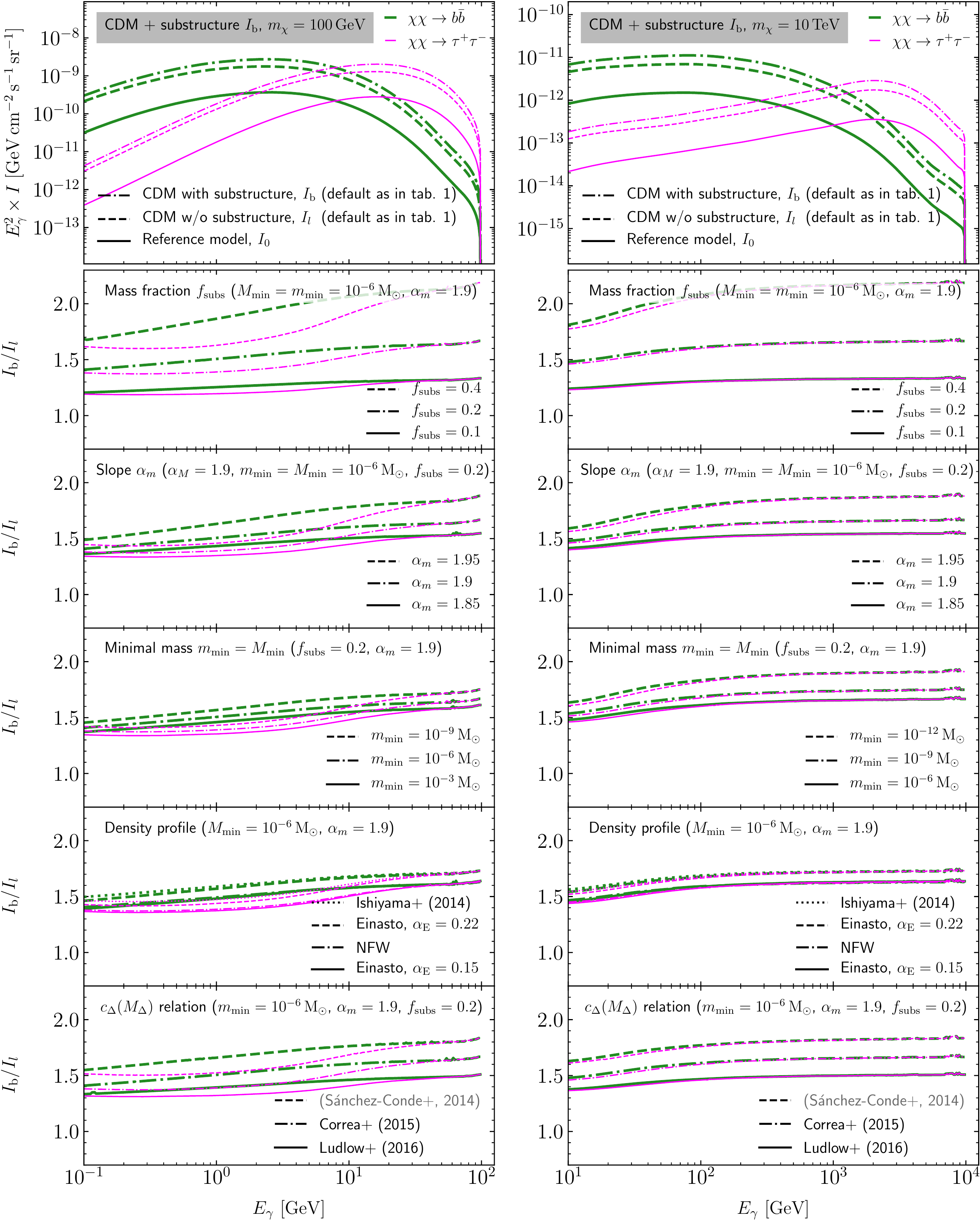}
\caption{
{\em Top row:} Intensity $I_{\rm b}$ (all halos with subhalos) for the same WIMP configurations as in \cref{fig:DiffFluxRatioHMFuncertainties}, where we also report $I_0$ ($M\geq10^{10}$~\Msol{}, no subhalos) and $I_l$ (all halos, no subhalos) for comparison purpose. {\em Remaining rows:} Ratios of $I_{\rm b}/I_l$ varying subhalo properties (\cref{tab:benchmarkmodel}). The bracketed model is excluded from the error budget. See discussion in \cref{subsubsec:substructure}.}
\label{fig:DiffFluxRatioSubsUncertainties}
\end{figure}

As in the previous sections, we discuss in turn the various ingredients and their impact on the overall boost. To start with, the top panel of \cref{fig:DiffFluxRatioSubsUncertainties} shows, for the reference subhalo configuration (see \cref{tab:benchmarkmodel}), the intensity accounting for the boost, $I_{\rm b}$, compared to the no-boost case $I_l$. The boost is not even a factor two, which is understood as follows. First, the overall boost is the integration of the individual boosts of all host halos over redshifts and masses. As illustrated in \cref{sec:intensity_multiplier_bricks} and \cref{fig:Fluxmultiplier_luminosities}, the trend is to have a decreasing boost with decreasing $M_{\rm host}$ and increasing $z$. At $z=0$, the boosts we obtain are consistent with the results of \cite{2014MNRAS.442.2271S}; the decrease with $z$ is related to the decreasing concentration of halos with redshift. Second, as seen in the previous section, the intensity from field halos in the mass range below $\unit[10^{10}]{\Msol}$ dominates the high-mass range contribution ($I_l$ compared to $I_0$ in the top panel). In this mass range, mild to no boost is expected, so that this combines to give an overall very mild boost. In more details, the subhalos parameters play as follows.

\paragraph{Mass fraction $f_{\rm subs}$ in subhalos.} Studying the mass and redshift evolution of $f_{\rm subs}$ from several numerical simulations, \cite{2017MNRAS.472..657J} found that $f_{\rm subs}$ grows from 6\% for $\unit[10^{10}]{\Msol}$ halos to 17\% for $\unit[10^{15}]{\Msol}$ halos, and that $f_{\rm subs}$ increases for more recently formed halos of a given mass. This range may be even larger, with a value of 2\% obtained for galaxy halos in \cite{2008MNRAS.391.1685S} and 30\% for galaxy cluster halos in \cite{2012MNRAS.425.2169G}. In their DGRB calculation, the \fermi{} collaboration even argue for a value $f_{\rm subs}=0.45$ \cite{2015JCAP...09..008T}. To encompass the range of possible values for the full range of halo masses and redshifts, we vary $f_{\rm subs}$ from 0.1 to 0.4 (0.2 is our default value) in the second panel of \cref{fig:DiffFluxRatioSubsUncertainties}. This leads to a variation of the ratio $I_{\rm b}/I_l$ in $\sim1.2-2$; smaller fractions would converge to no boost.

\paragraph{Slope $\alpha_{m}$ of subhalo mass function.}
We parametrise the subhalo mass function as a pure power-law, $\dd n/\dd m\propto m^{-\alpham}$, between $m_{\rm min}$ and $m_{\rm max} = 10^{-2}\times M_{\rm host}$ (using $10^{-1}\times M_{\rm host}$ lowers the result by 5\% only). The slope \alpham{} in numerical simulations is typically found in $1.85\lesssim\alpham\lesssim 1.95$ \cite{2008MNRAS.391.1685S}. Note that the slope reported in some recent simulations indicates $\alpham\sim 1.75$ \cite{2016MNRAS.462..893R,2017MNRAS.469.4157C}, but this assumes a power-law with an exponential cut-off, which translates into $\alpham\sim 1.9$ for a pure power law. A recent meta-analysis of the subhalo population from many simulations finds a good agreement between all the simulations, with $\alpham\sim 1.85$ \cite{2016MNRAS.458.2870V} despite the use of different halo finders. In the third panel of \cref{fig:DiffFluxRatioSubsUncertainties}, we vary this slope from $1.85$ to $1.95$ (with $1.9$ the default value) and find that this has almost no impact on the result.

\paragraph{Minimal subhalo mass, $m_{\rm min}$.}
Assuming that all subhalos from the free streaming scale survive in their host halos, we match the minimal mass of the subhalos to that of the minimal mass taken for field halos (see \cref{subsec:lowmass_contrib}). The fourth row of \cref{fig:DiffFluxRatioSubsUncertainties} shows a small scatter of $\lesssim15\%$ between the various mass cut-offs assumed. This indicates that unless all subhalos are destroyed up to masses much larger than $m_{\rm min}=\unit[10^{-3}]{\Msol}$, the overall boost of the extragalactic signal is not very sensitive to the exact minimal mass of the subhalos.

\paragraph{Density profiles of subhalos.} As for field halos, we test the impact of various density profiles for subhalos, including two extreme values for the Einasto slope, using an NFW profile, or the Ishiyama model with cuspier profiles for micro-halos. The result on the boost is shown in the next-to-last row of \cref{fig:DiffFluxRatioSubsUncertainties}, with $I_{\rm b}/I_0$ varying in $\sim 1.4-1.7$.

\paragraph{Mass-concentration-redshift relation (and subhalo spatial distribution).}
Micro-halos can be tidally disrupted or partially stripped preferentially in the inner parts of galaxies \cite{2015MNRAS.447.1353M,2017MNRAS.471.1709G,2017PhRvD..95f3003S}. This impacts the concentration of subhalos compared to field halos \cite{2015PhRvD..92l3508B,2017MNRAS.466.4974M}, making surviving subhalos more concentrated in the inner parts of the halos. While this has an important effect for the annihilation signal in the Galaxy and dark clump detection (e.g., \cite{2016JCAP...09..047H}), this mostly affects the inner parts which do not contribute much to the total annihilation signal. Moreover, beside the modelling of \cite{2017MNRAS.466.4974M} for Galactic subhalos at $z=0$, a complete study of the subhalo mass-concentration relation with mass and redshift is lacking. For this reason, we assume the same concentration for field and subhalos, and the last row of \cref{fig:DiffFluxRatioSubsUncertainties} shows a scatter of $\pm 15\%$ from the different models used. For the sake of completeness, we also study the impact of the subhalo spatial distribution: antibiased distributions have been obtained in DM simulations of galaxies \cite{2008MNRAS.391.1685S} and galaxy clusters \cite{2012MNRAS.425.2169G}. Using an unbiased distribution, i.e., the same profile as for the host, or the two above biased distributions have a marginal effect ($<1\%$) on $I_{\rm b}/I_l$.


\section{Discussion}
\label{sec:discussion}
The exotic \gr{} intensity from DM annihilation has been previously computed by various authors and we provide comparisons to some of these earlier studies in \cref{fig:DiffFluxComparison}. Our results are given by the green lines as follows: (i) the 'robust' lower bound $I_0$ described in \cref{subsec:benchmark_model} is plotted in solid green lines, (ii) the calculation extrapolated to the minimum field halo mass $I_l$ (\cref{subsubsec:small_mass_divergence}) is given by the green dashed lines and (iii) finally, the intensity estimation including boost from subhalos $I_{\rm b}$ (\cref{subsubsec:substructure}) is shown as dotted-dashed lines. The green-shaded bands correspond to the cumulative uncertainties summarised in \cref{tab:benchmarkmodel} on the lower bound from high-mass halos only, $I_0$, and the CDM cases including small-scale structure. The diffuse \gr{} background measured by \fermi{} \cite{2015ApJ...799...86A} and the systematic uncertainty band from the foreground modelling are given in blue.
\subsection{Comparison to other works}
\begin{figure}[t!]
  \begin{center}
    \includegraphics[width=0.79\textwidth]{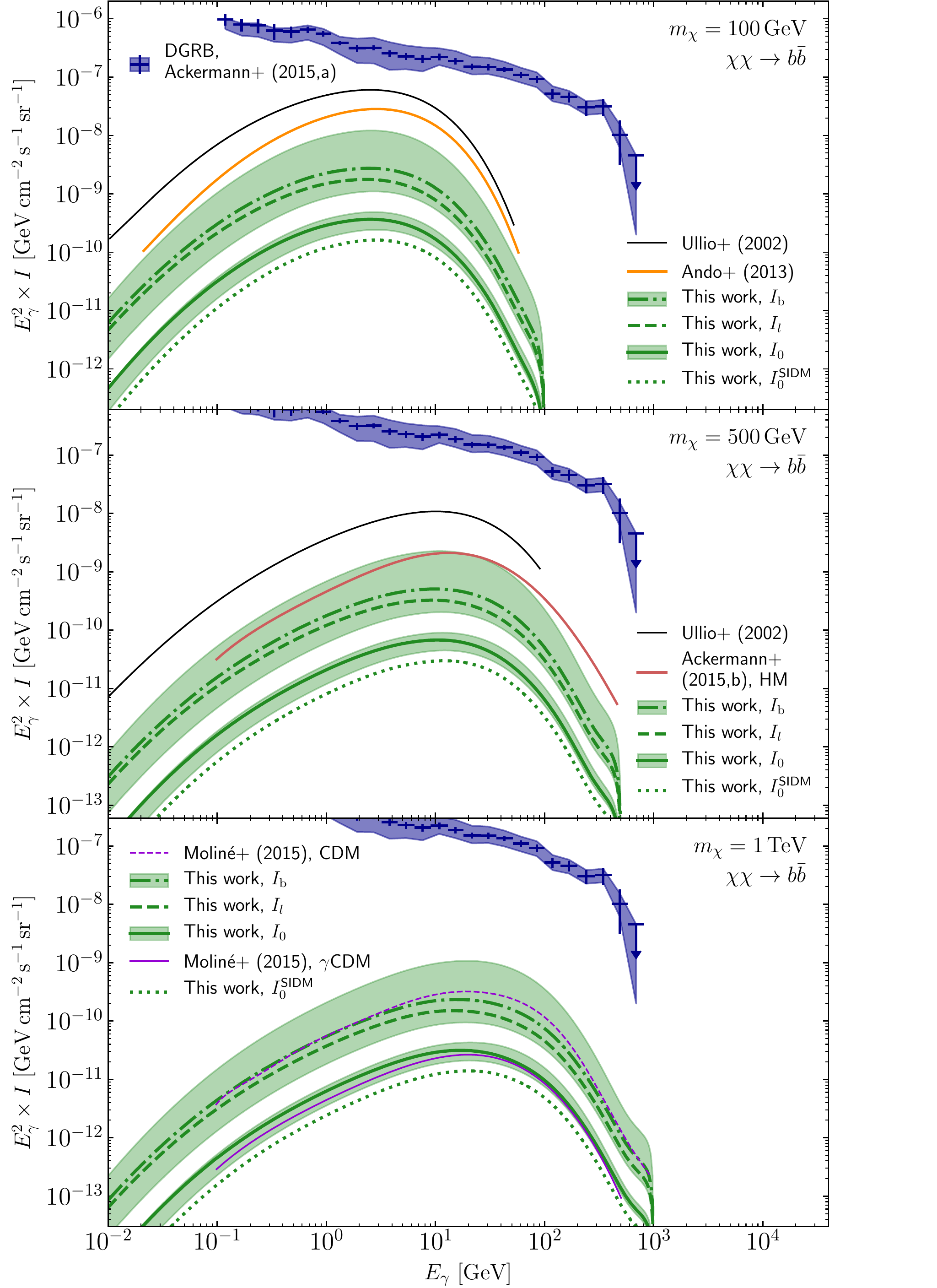}
  \caption{
Extragalactic DM annihilation intensity for $\mchi$=100 (top), 500 (middle), and $10^3$~GeV, for pure $\chi\chi\rightarrow b\bar{b}$ annihilation and $\sigmav=\unit[3\times 10^{-26}]{cm^3\,s^{-1}}$: our results for collisionless CDM are shown for high-mass halos only ($M\geq10^{10}$~\Msol{}, solid green curves), the intensity of all halos without (dashed green curves) or with (dot-dashed green curves) substructures. The green uncertainty bands are obtained from extreme choices of the ingredients in \cref{tab:benchmarkmodel}. The total intensity is compared to selected previous works (black  \cite{2002PhRvD..66l3502U}, orange  \cite{2013PhRvD..87l3539A}, red  \cite{2015JCAP...09..008T}, and violet \cite{2016JCAP...08..069M} curves). The intensity for the SIDM scenario, as discussed in \cref{subsec:SIDM}, is shown as a dotted green curve. We also report the DGRB intensity as measured by \fermi{} \cite{2015ApJ...799...86A} (foreground model A), where the blue shaded band denotes their systematic uncertainty due to the Galactic foreground modelling.
}
  \label{fig:DiffFluxComparison}
  \end{center}
\end{figure}

Except when comparing to the recent estimation of Molin\'e et al. (2016) \cite{2016JCAP...08..069M} (lower panel, violet), the estimated \gr{} intensity derived in this work is lower than the previously published results of Ullio et al. (2002, \cite{2002PhRvD..66l3502U}, black solid lines), Ando et al. (2013, \cite{2013PhRvD..87l3539A}, orange solid line) and Ackermann et al. (2015, \cite{2015JCAP...09..008T}, red solid line). Among the many studies led on the topic, we selected these four analyses for comparison as, as a whole, they are representative of the evolution of the exotic extragalactic \gr{} emission calculations in the last fifteen years. The origins of the differences with our estimation depend on the study under scrutiny:
\begin{itemize}
\item In \cite{2002PhRvD..66l3502U}, the authors used a Moore parametrisation for the DM halo profiles and a mass concentration according to Bullock et al. (2001, \cite{2001MNRAS.321..559B}), which will both increase the intensity (black solid line in top and middle panels). The Moore DM profile, with an inner logarithmic slope of 1.5, diverges at small radii and can yield very large values of the luminosity in \cref{eq:luminosity}.
\item While somewhat lower than that of \cite{2002PhRvD..66l3502U}, the prediction of \cite{2013PhRvD..87l3539A} (orange line in top panel) is still much higher than our estimation. This is understood as they used an effective subhalo boost model fitted to the results of \cite{2012MNRAS.425.2169G}, typically $\sim 20$ at halo masses of $M=\unit[10^{10}]{\Msol}$, $\sim 100$ at $\unit[10^{12}]{\Msol}$, and $\sim 1500$ at $\unit[10^{15}]{\Msol}$. Adopting such values  would indeed bring the total intensity $I_\mathrm{b}$ to the level of their result. Since then, such large boosts have been excluded by several authors \cite{2012MNRAS.425..477N,2014MNRAS.442.2271S,2017MNRAS.466.4974M}, including this work (see \cref{fig:Fluxmultiplier_luminosities}), when using physically-motivated mass-concentration relations at low masses.
\item We also compare our result to that of the \fermi{} collaboration \cite{2015JCAP...09..008T} (red curve in middle panel), whose estimate is consistent with the upper limit of our uncertainty band. Conversely to the previous cases, the difference with our reference is not due to a single major feature of their modelling, but from small differences in the extrapolation to lower mass (e.g., steeper slope of the subhalo mass distribution, and possibly of the mass function extrapolation, larger subhalo mass fraction) which yield, when combined, the factor $\sim 5$ between $I_{\rm b}$ and their result in the middle panel of \cref{fig:DiffFluxComparison}. The authors also explored the uncertainties using the non-linear power spectrum instead of the halo model approach we used, noting that the many dependencies of the halo model render this task more difficult. It is interesting to note that the uncertainty band we find by varying all the various ingredients in the halo model is similar, although twice smaller, to what \cite{2015JCAP...09..008T} find using the power spectrum approach.
\end{itemize}

\subsection{A word on IDM (SIDM/$\gamma$CDM) models}
\label{subsec:SIDM}
So far, we have remained in the framework of collisionless, cold DM. Beyond this paradigm, the concept of  interacting DM (IDM) has raised the attention of the community in the last years as a possible solution to various observational tensions on subgalactic scales between collisionless CDM and observations. These include the observation of pronounced DM halo cores from dwarf galaxy to cluster sizes \cite{1994Natur.370..629M,2016PhRvL.116d1302K}, the problems of the diversity of rotation curves \cite{2015MNRAS.452.3650O} and missing satellites \cite{2010arXiv1009.4505B}, and the ``too-big-to-fail'' problem \cite{2011MNRAS.415L..40B}. In contrast to warm DM (WDM) made of  sub-MeV particles as a possible solution to CDM small-scale issues, IDM may solve these problems while at the same time preserving the remarkable successes of CDM. Most importantly in our context and in contrast to WDM, IDM could still consist of cold and heavy DM particles annihilating into high-energy \grs{}.

In a self-interacting DM (SIDM) scenario, the DM particle has a non-negligible cross section, $\sigma_{\rm el}$, of mutual weak elastic scatterings \cite{2017arXiv170502358T}. A scattering cross section $\sigma_{\rm el}$ in the order of magnitude of $\sigma_{\rm el}/\mchi\sim\unit[0.1]{cm^2/g}$  would be still in agreement with observation, e.g., the non-observation of self-interactions in cluster mergers \cite{2008ApJ...679.1173R} or halo-shape constraints \cite{2013MNRAS.430..105P} (see also \cite{2017arXiv170502358T} for further references). At the same time, such a self-interaction would be large enough to repeal the diversity and cusp-vs-core problems by increasing the scatter in concentrations and thermalising the inner halos, removing the inner density cusps of Einasto or NFW halos. On the other hand, it has been shown that a velocity-independent $\sigma_{\rm el}/\mchi\sim\unit[0.1]{cm^2/g}$ is too low to solve the missing-mass problem (i.e., not able to significantly reduce the number density of small-mass halos) \cite{2013MNRAS.431L..20Z}. Therefore, elastic scattering between the DM and photons (``$\gamma$CDM'') or neutrinos has been proposed as a more general class of interacting DM (IDM) \cite{2014MNRAS.445L..31B}. It has been shown that $\gamma$CDM scatterings with cross sections on the electroweak scale likewise reduce the core concentrations of MW-like DM halos and can also efficiently suppress structures at the dwarf-galaxy scale, $M\lesssim \unit[10^{10}]{\Msol}$ \cite{2015MNRAS.449.3587S,2016JCAP...08..069M}. As shown in \cref{fig:DiffFluxRatioHMFextrapolation}, \cite{2016JCAP...08..069M} find, for a $\gamma$CDM scattering with $\sigma_{\rm el}/\mchi = \unit[7.5\times10^{-10}]{cm^2/g} \sim \unit[1]{nb/GeV} $,\footnote{See \cite{2014MNRAS.445L..31B} for a detailed motivation of the chosen value.} the halo mass function is suppressed at just around the scale of our reference model, $I_0$, with a half-mode mass of  $M_{\rm hm}=\unit[4.3\times10^{9}]{h^{-1}\,\Msol}$.

We shortly explore the impact of thermalised halo cores in such models, as such a flattening of the central densities may significantly affect the \gr{} signal from DM annihilation. To this purpose, we remain in the framework of SIDM and assume a SIDM scattering cross section of $\sigma_{\rm el}/\mchi\sim\unit[1]{cm^2/g}$. Although in tension with several constraints, this allows to conservatively estimate a maximal reduction of the \gr{} intensity from cored inner DM halos. For such a SIDM cross section, the authors of \cite{2013MNRAS.430...81R} provide a simple scaling relation, to which we add a redshift evolution of the comoving core density according to
\beq
\rho_{\rm core}(M_{\rm vir},\,z) = \frac{0.029}{(1+z)^3}\,\unit{\frac{\Msol}{pc^3}} \times \left(\frac{M_{\rm vir}}{\unit[10^{10}]{\Msol}}\right)^{-0.19}\,,
\label{eq:rhocore_sidm}
\eeq
for a Burkert halo profile \cite{1995ApJ...447L..25B}
\beq
\rho^{\rm Burkert}(r) = \frac{\rho_{\rm core}\,r_{\rm core}^3}{(r + r_{\rm core})(r^2 + r_{\rm core}^2)}\,.
\eeq
This relation corresponds to core radii of $r_{\rm core}(\unit[10^{10}]{\Msol}) = \unit[2.22]{kpc}$ and $r_{\rm core}(\unit[10^{15}]{\Msol}) = \unit[244]{kpc}$ at $z=0$, respectively. At higher redshifts, \cref{eq:rhocore_sidm} results in halo concentrations $c_{\rm vir}(z)\sim c_{\rm vir}(0)/(1+z)$ which we have found to provide lower intensities compared to more recent prescriptions (see \cref{subsec:benchmark_model}). We retain $M_{\min}= \unit[10^{10}]{\Msol}$ for this assessment of DM self-interactions, although  smaller structure may be still present for SIDM. We obtain that a core condition according to \cref{eq:rhocore_sidm} reduces the halo luminosities by approximately a factor 2  in the range $\unit[10^{10}]{\Msol}\lesssim M \lesssim \unit[10^{15}]{\Msol}$ at $z=0$ and  more at higher redshifts. As shown in \cref{fig:DiffFluxComparison} (green dotted curves), the intensity $I_0^{\text{SIDM}}$ from halos with SIDM cores is a factor of $2-5$ (depending on $E_\gamma$)  lower than the corresponding reference intensity, $I_0$.

\section{Conclusions}
\label{sec:conclusion}

In this work, we have reassessed the isotropic emission of \grs{} for extragalactic annihilations of WIMPs in a clumpy $\Lambda$CDM Universe. We have also ranked the various sources of uncertainties, in order to identify which ingredients dominate the error budget. The results are based on the latest available knowledge about structure formation in a \planck{} cosmology.

For CDM, we first calculate the intensity $I_0$ from DM halos with $M\geq 10^{10}$~\Msol{}, a mass range where no extrapolation is required. This contribution constitutes a lower bound on the \gr{} emission from extragalactic DM and we find $I_0$ can be robustly estimated within a factor 2. Accounting for the population of halos and subhalos down to the smallest masses, the total intensity, $I_{\rm b}$, is a factor $\sim 10$ larger than $I_0$ and bracketed by a one order of magnitude uncertainty band. Recent estimations of $I_{\rm b}$ (including our own) tend to find lower values than earlier studies; this trend can be traced to differences in the mass function extrapolation and  to the recent predictions of smaller mass concentrations in small-scale halos. In particular, we conclude in a marginal boost  ($\sim 1.5$) of the signal by halo substructures.

The fiducial spectrum of the intensity $I_{\rm b}$ obtained in this work is a factor 5 smaller (with slightly smaller uncertainties) than that derived in the \fermi{} analysis \cite{2015JCAP...09..008T}. Taken at face value, this would relax the corresponding DM exclusion limit by the same amount, making the DGRB an even less competitive target w.r.t. dwarf spheroidal galaxies.

For SIDM with $\sigma_{\rm el}/\mchi\sim\unit[1]{cm^2/g}$, we find that the signal of our lower bound $I_0$ is further reduced by a factor $\sim 3$ compared to the collisionless case. As we generally expect fewer or even no boost from small structures in interacting DM scenarios, the extragalactic signal may remain three (for $\mchi=\unit[100]{GeV}$) to four (for $\mchi=\unit[1]{TeV}$) orders of magnitudes below the measured DGRB intensity for a canonical annihilation cross section of $\sigmav=\unit[3\times 10^{-26}]{cm^3\,s^{-1}}$. This indicates that if a \gr{}  signal from DM self-annihilations is seen in dwarf spheroidal galaxies, whose signal is not expected to change much between interacting and collisionless DM scenarios, a non-observation of a corresponding signal in the DGRB could be used to probe properties of elastic DM interactions in SIDM and $\gamma$CDM/$\nu$CDM models.

All the results presented in this work have been performed with a soon to be released version of the \clumpy{} code. Using the latter, the calculation could be easily repeated to assess \textit{neutrino} signals from extragalactic DM, for which the various final state neutrino spectra are already included in the code.

\acknowledgments

We thank J. Lesgourgues for helpful input on the \texttt{CLASS} power spectrum. This work has been supported by the Research Training Group 1504, ``Mass, Spectrum, Symmetry'', of the German Research Foundation (DFG), by the ``Investissements d'avenir, Labex ENIGMASS", and by the French ANR, Project DMAstro-LHC, ANR-12-BS05-0006.


\appendix

\section{Mass conversion between different $\Delta$ definitions}
\label{sec:mdeltaconversion}
Let us assume a generic halo density profile, $\rho_{\mathrm{halo}}(r;\,\rho_{-2},\,r_{-2},\,\vec{\alpha})$ with a given set of $\vec{\alpha}$: for instance, $\vec{\alpha}=\alpha_E$ for an Einasto or $\vec{\alpha}=(\alpha,\beta,\gamma)=(1,3,1)$ for an NFW profile. We want to calculate its  mass $M_{\Delta_1}$ for a corresponding $\Delta_1$, given a mass-concentration relation w.r.t. $\Delta_{\rm ref}(\neq \Delta_1)$,
\beq
c_{\Delta_{\rm ref}}(M_{\Delta_{\rm ref}})=\frac{R_{\Delta_{\rm
ref}}}{r_{-2}}(M_{\Delta_{\rm ref}})\,.
\label{eq:cdeltaref}
\eeq
Provided a density profile of the generic form
\beq
\rho_{\mathrm{halo}} = \rho_{-2}\times
\widetilde{\rho}(r/r_{-2};\,\vec{\alpha})= \rho_{-2}\times
\widetilde{\rho}(x;\,\vec{\alpha})\,,
\eeq
one can solve the implicit equation
\beq
 \frac{M_{\Delta_i}}{M_{\Delta_{\rm ref}}} \times \frac{\int\limits_0^{c_{\rm ref}}x^2\, \widetilde{\rho}(x;\,\vec{\alpha})\,\dd x}{\int\limits_0^{c'_{\rm ref}}x^2\, \widetilde{\rho}(x;\,\vec{\alpha})\,\dd x} = 1\;
\label{eq:m1m2_3}
\eeq
for $M_{\Delta_{\rm ref}}$, given $\Delta_i = \Delta_1$, and with the expressions
\begin{align}
c_{\rm ref} &:= c_{\Delta_{\rm ref}}\left(M_{\Delta_{\rm
ref}},z\right)\,,\\
c'_{\rm ref} &:= c_{\Delta_{\rm ref}}\left(M_{\Delta_{\rm ref}},z\right)\times
\left(\frac{M_{\Delta_i}}{M_{\Delta_{\rm ref}}}\times \frac{\Delta_{\rm ref}(z)}{\Delta_i(z)} \right)^{\frac{1}{3}}\,.
\end{align}
Now, knowing $M_{\Delta_{\rm ref}}$, \cref{eq:m1m2_3} can be
solved a second time for $\Delta_i = \Delta_2$ to directly convert the halo mass from an arbitrary
$\Delta_1$ to $ \Delta_2$ without the need to determine $\rho_{-2}$ and $r_{-2}$ of the halo. Although computationally more expensive, this rigorous algorithm is more flexible and precise for arbitrary choices of mass-concentration relations and halo profiles than the approximate translation recipe from \cite{2010MNRAS.404..502G} which we have used in an earlier work \cite{2016JCAP...09..047H}.


\section{Behaviour of the various terms in the intensity multiplier}
\label{sec:intensity_multiplier_bricks}

In this appendix, we provide more details about the ingredients we use to compute the intensity multiplier, \cref{eq:delta_dm_annihil}, and intermediate results.

\begin{figure}[t]
\centering
\includegraphics[width=\textwidth]{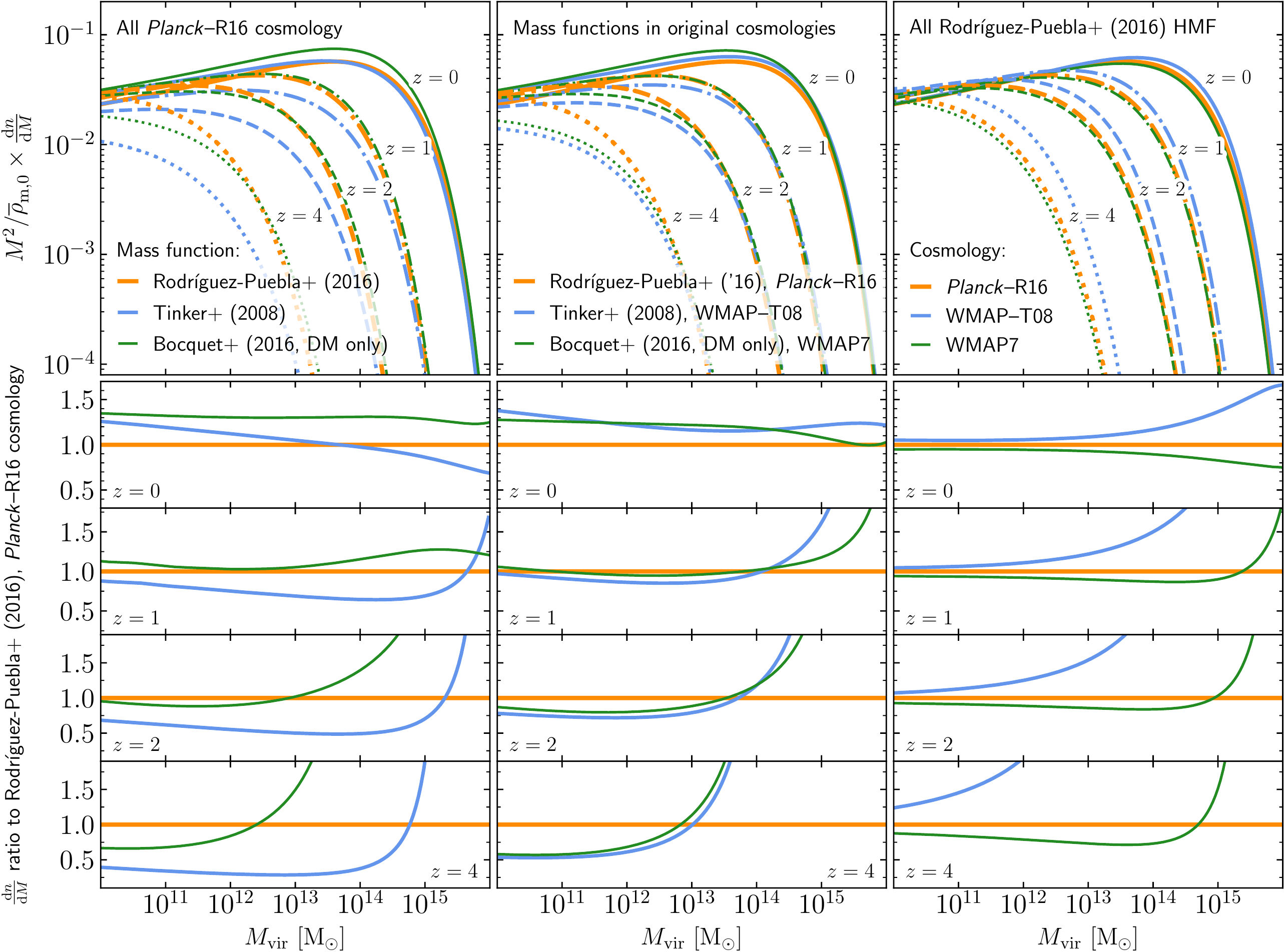}
\caption{
{\it Left panels:} Comparison of different mass functions in the same cosmology of  \planck{}--R16 \cite{2016MNRAS.462..893R}.
{\it Central panels:}  Comparison of different mass functions in the cosmology of the simulations they were derived from.
{\it Right panels:} Comparison of the mass function from \cite{2016MNRAS.462..893R} rescaled to different cosmologies. Note that $\overline{\varrho}_{\rm m, 0}$ differs between the cosmologies, while the ratios are given with respect to $\dd n/\dd M$.
}
\label{fig:HMF_comparison_threeInRow}
\end{figure}

\paragraph{Halo mass function/multiplicity function.} In this work, we have used various descriptions of the multiplicity function, $f(\sigma,\,z)$ (\cref{eq:halomassfunction}), in different cosmologies. For validation purpose of the discussion in \cref{subsec:benchmark_model}, we show in \cref{fig:HMF_comparison_threeInRow} the underlying halo mass functions rescaled to $\Delta=\Delta_{\rm vir}$: this emphasises the impact of changing the multiplicity function in a given cosmology (right panel of \cref{fig:HMF_comparison_threeInRow}) or changing the cosmology underlying the linear matter power spectrum to compute $\sigma(M)$, \cref{eq:sigma2}, (left panel). Note the remarkable concordance of the mass functions from T08 \cite{2008ApJ...688..709T} and B16 \cite{2016MNRAS.456.2361B} in their ``original'' cosmologies and in the $\Delta_{\rm vir}$ prescription (blue and green curves in the central panel).

\begin{figure}[t]
\centering
\includegraphics[width=0.59\textwidth]{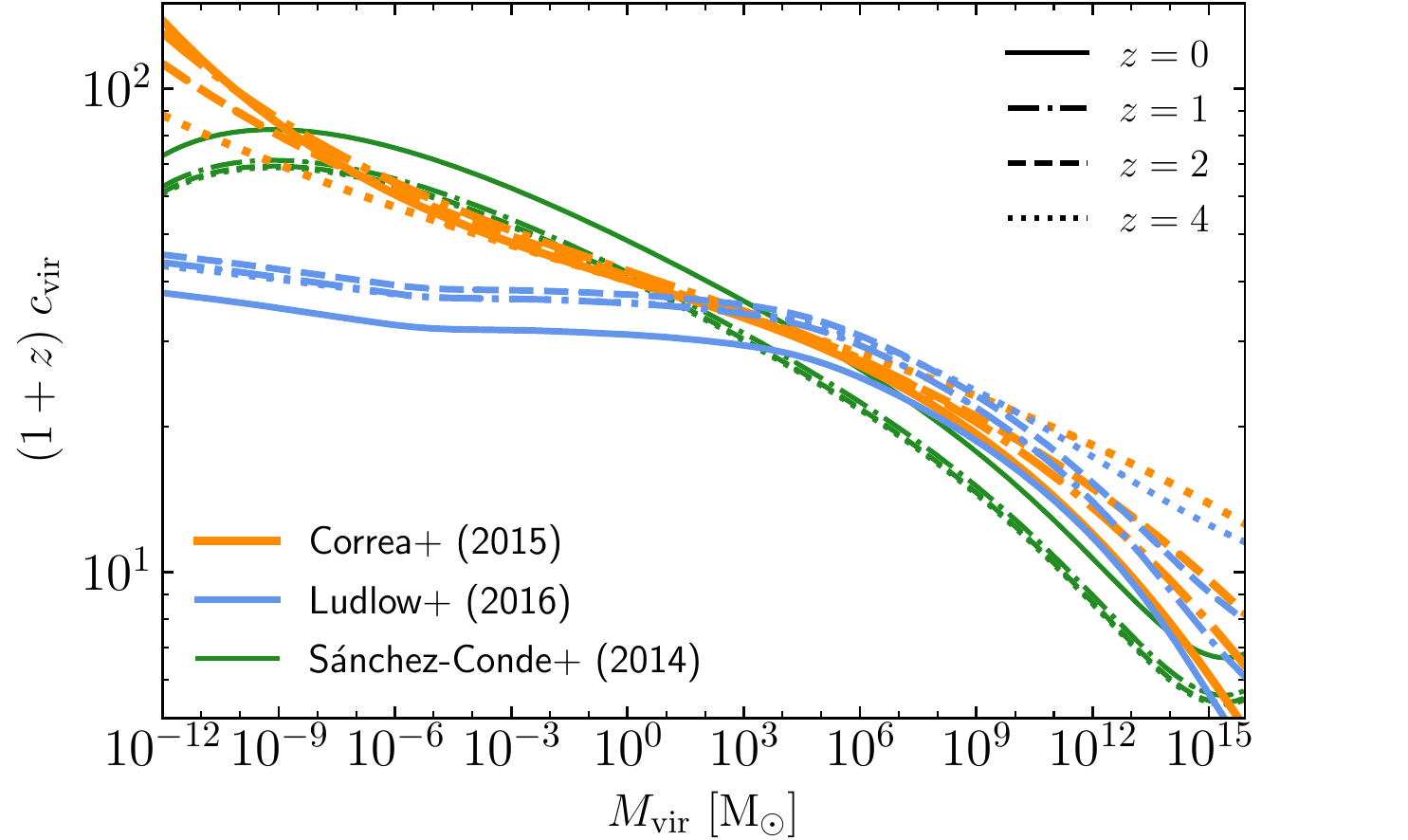}
\caption{
Mass-concentration relations considered in this work w.r.t the overdensity $\Delta_{\rm vir}$, \cref{eq:bryannorman}. The rescaling to $\Delta_{\rm vir}$ from the original prescriptions is done according to \cref{sec:mdeltaconversion} and using an Einasto profile. Note that the model L16 \cite{2016MNRAS.460.1214L} is based on the density variance $\sigma$, which is computed here using \cref{eq:sigma2} and not from the analytical approximation provided in L16.}
\label{fig:Concentrations_1+z}
\end{figure}

\paragraph{Mass-concentration-redshift parametrisation $c(M,z)$.}
\Cref{fig:Concentrations_1+z} displays the different models of halo concentrations compared in this work for an overdensity definition of $\Delta_{\rm vir}$ according to \cref{eq:bryannorman}. The original concentration relations for all these three models are given w.r.t $\Delta_{\rm c} = 200$. In this figure, we used the algorithm described in the previous \cref{sec:mdeltaconversion} to translate the concentration relations to different choices for $\Delta$, assuming an Einasto halo density profile with $\alpha_{\rm E}= 0.17$.
This rescaling between different definitions for $\Delta$ explains why the green curves for the model S14 are not overlapping when displaying $(1+z)\,c_{\rm vir}$: While  the original prescription from  \cite{2014MNRAS.442.2271S} gives $c_{200,\rm c}(M_{200,\rm c},z)  =  c_{200,\rm c}(M_{200,\rm c},z=0)/(1+z)$, our recipe results in a $c_{\rm vir}(M_{\rm vir},z)$ decreasing even stronger with redshift.\footnote{Note that $\Delta_{\rm vir}(z=0)\approx 100$. For $z\gtrsim 1$, $\Delta_{\rm vir}(z)$ converges towards $178<\Delta_{\rm c}=200$, as $\Omega_{\rm m}\rightarrow 1$ in the matter dominated era.} In turn, the model L16 \cite{2016MNRAS.460.1214L} (blue curves) shows a weaker scaling than $\propto(1+z)^{-1}$ in both the original description and after translation to $\Delta_{\rm vir}$. Our default model C15 \cite{2015MNRAS.452.1217C} (orange curves) remarkably well scales  $\propto(1+z)^{-1}$ relative to  $\Delta_{\rm vir}$ over several mass decades, $\unit[10^{-6}]{\Msol}\lesssim M_{\rm vir}\lesssim \unit[10^{8}]{\Msol}$.

\begin{figure}[t]
\centering
\includegraphics[width=\textwidth]{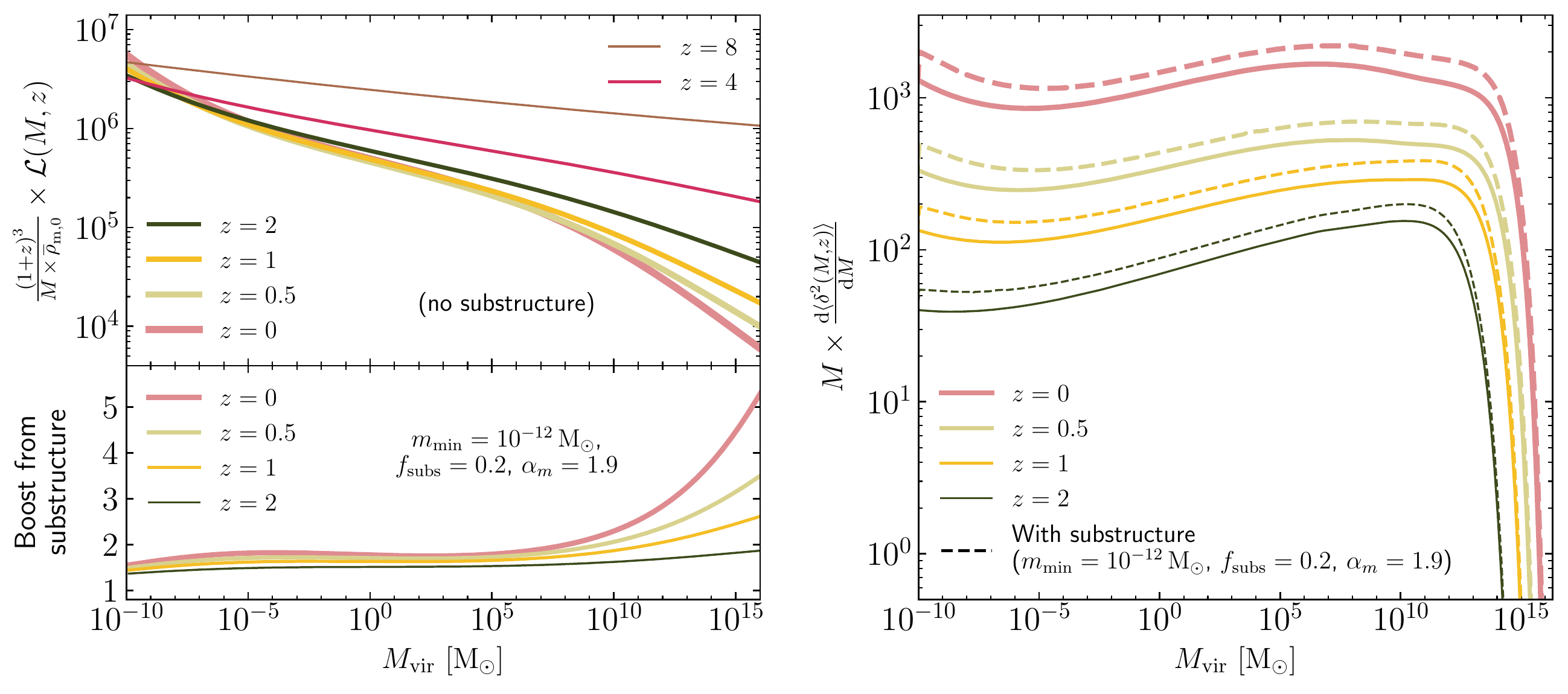}
\caption{
Contributors to the intensity multiplier, $\langle\delta^2\rangle$, \cref{eq:delta_dm_annihil}, from different halo mass decades. {\it Upper left panel:} Comoving one-halo luminosities at different redshifts, \cref{eq:luminosity}, in the reference model and without halo substructures. {\it Lower left panel:} Boost factor to the one-halo luminosities at different redshifts due one level of halo substructures and the  $c(M,z)$ model C15 \cite{2015MNRAS.452.1217C}. It can be seen that the substructure boost decreases with redshift and host halo mass. {\it Right panel:} Integrand of the mass integral of \cref{eq:delta_dm_annihil} at $z\leq 2$ with and without substructure boost. Note that for display purpose, contrarily to the left, we do not multiply $\dd \langle\delta^2\rangle/\dd \log M$ with a factor $(1+z)^3$ here.}
\label{fig:Fluxmultiplier_luminosities}
\end{figure}
\begin{figure}[t]
\begin{flushright}
\includegraphics[width=0.985\textwidth]{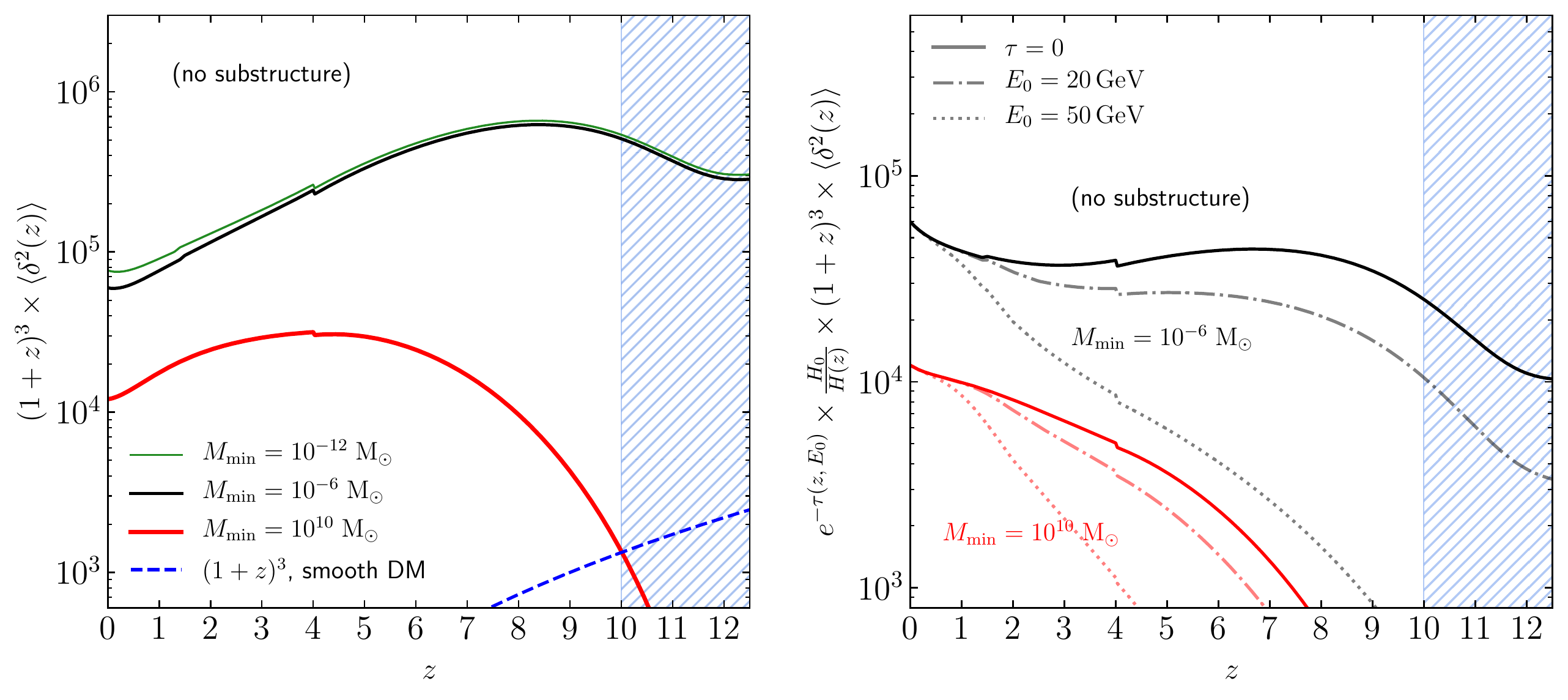}
\end{flushright}
\caption{
Redshift dependence of  the intensity multiplier, $\langle\delta^2\rangle$, \cref{eq:delta_dm_annihil}, for our reference model (red curves) and the default collisionless CDM case without substructures (black curves). On the left, we show the comoving multiplier (we additionally display the impact of adopting $M_{\rm min}= \unit[10^{-12}]{\Msol}$, green curve), whereas on the right, the multiplier is additionally multiplied by the volume element and different EBL absorption factors.  The step in the curves originates from the $c_{\Delta}$ model by \cite{2015MNRAS.452.1217C}, which connects two functional forms at $z=4$. See text for further details.
}
\label{fig:Fluxmultiplier_integrated}
\end{figure}

\paragraph{One-halo luminosities, $\mathcal{L}(M,z)$, and substructure boost.}
\Cref{fig:Fluxmultiplier_luminosities} (upper left) shows the DM one-halo luminosities $\mathcal{L}$, \cref{eq:luminosity}, for our default $c_{\Delta}(M_{\Delta})$ model C15 \cite{2015MNRAS.452.1217C}. Displaying $(1+z)^3\times \mathcal{L}$,
it can be seen that the higher mean density of the early Universe|and by this, a higher annihilation rate|overcompensates the less concentrated halos at earlier epochs, and the comoving luminosity increases with redshift. This finally prevents convergence of the intensity multiplier over the considered redshift range in a collisionless CDM paradigm, as illustrated in the later \cref{fig:Fluxmultiplier_integrated}. The lower panel of \cref{fig:Fluxmultiplier_luminosities} shows the ratio of the emission with and without substructures, commonly referred to as the substructure boost. For the concentration model from \cite{2015MNRAS.452.1217C}, the emission is only moderately boosted by substructures by less than a factor five at galaxy cluster masses at $z=0$, and even less for smaller masses and higher redshifts.

\paragraph{Intensity multiplier, $\langle\delta^2\rangle$.} In \cref{fig:Fluxmultiplier_luminosities} (right) we show the contribution per mass decade to the intensity multiplier, \cref{eq:delta_dm_annihil}. Here, the one-halo luminosities are multiplied with the halo number density $\dd n/\dd M$, which is extrapolated to the micro-halo scale according to a power-law extrapolation, \cref{eq:pl_hmf_Extrapolation} with $\alphaM=1.9$. We show $\dd \langle\delta^2\rangle/\dd \log M$ without (solid lines) and with (dashed lines) emission boost from substructures. From this depiction it becomes evident that a cut-off of the mass function at some minimal mass $M_{\rm min}$ is compulsory for the integral of \cref{eq:delta_dm_annihil} not to diverge. Previously, it has been found that for optimistic assumptions about the substructure boost, the intensity multiplier is dominated by galaxy and cluster-size halos, $M\gtrsim \unit[10^{10}]{\Msol}$ \cite{2013PhRvD..87l3539A}. However, this dominance is not present for our more moderate assumption of the concentration model C15, as shown in \cref{fig:Fluxmultiplier_luminosities} (left), for which all mass decades are similarly boosted (dashed lines).

\Cref{fig:Fluxmultiplier_integrated} finally shows the full intensity multiplier as a function of the redshift. The left panel shows that in a collisionless CDM scenario (power-law extrapolation of the mass function with $\alphaM=1.9$ down to $M_{\rm min}= \unit[10^{-6}]{\Msol}$, black curve, resp. $M_{\rm min}= \unit[10^{-12}]{\Msol}$, green curve), the density variance remains $\langle\delta^2\rangle\approx \mathrm{Var}(\delta)\gtrsim 100$ until high redshifts, $z\gtrsim 10$. For our reference model, where we only consider structures and halos on comoving mass scales larger than $\unit[10^{10}]{\Msol}$, the intensity multiplier from halos becomes smaller than 1 at $z\approx 10$ (red curve on the left). Note that the emission from structureless DM, $\langle\delta^2\rangle\equiv 1$,  exceeds the emission from high-mass halos|our reference model|at redshifts $z\gtrsim 10$ (blue dashed curve on the left). On the right panel, we show the intensity multiplier multiplied with the volume element, expressed by the inverse Hubble constant, and the EBL attenuation factor. This  quantity is finally integrated in \cref{eq:mean_intensity} over the redshift to obtain the total extragalactic DM \gr{} intensity. However,  we only integrate the intensity up to $z_\mathrm{max}=10$, and the blue hatched areas in \cref{fig:Fluxmultiplier_integrated} indicate the redshift regime which we exclude in our calculations. The contribution to the intensity from $z>z_{\rm max}$ is marginal when considering only masses $M \geq \unit[10^{10}]{\Msol}$ (red curves) or in the presence of EBL absorption (dotted and dashed-dotted curves), however not necessarily in the presence of small-scale clustering and no absorption (black solid line on the right).

\bibliography{Huetten2017_JCAP}
\end{document}